\documentclass[12pt]{article}
\pdfoutput=1
\usepackage{tgpagella}
\usepackage[sc]{mathpazo}
\usepackage[top=3cm, bottom=3cm, inner=3.5cm, outer=3.5cm]{geometry} 
\usepackage{microtype}

\usepackage{amsfonts}
\usepackage{amssymb}
\usepackage{amsmath}
\usepackage{mathtools}
\usepackage{enumitem}
\usepackage{booktabs}
\usepackage{tabularx}
\usepackage{siunitx}
\usepackage[dvipsnames]{xcolor}
\usepackage[linktoc=all, pdfencoding=auto, psdextra]{hyperref}
\hypersetup{
	colorlinks,
	citecolor=MidnightBlue,
	filecolor=black,
	linkcolor=MidnightBlue,
	urlcolor=black	
}
\usepackage{csquotes}
\usepackage{graphicx}
\usepackage[font=small,labelfont=bf]{caption}
\usepackage{tikz}
\usepackage{calc}
\usetikzlibrary{arrows.meta, positioning}
\usepackage{pgfplots}
\usepackage{pgfplotstable}
\pgfplotsset{compat=1.18}
\usepgfplotslibrary{groupplots, colorbrewer}
\pgfkeys{/pgf/number format/.cd,1000 sep={\,}}
\usepackage[standard, amsmath, thmmarks]{ntheorem}
\usepackage[noabbrev]{cleveref}
\usepackage[
backend=biber,
style=numeric-comp,
urldate=long,
sortcites=true,
sorting=nyt,
url=true,
isbn=false,
doi=true
]{biblatex}
\bibliography{../../library.bib}
\usepackage{xurl}

\theoremseparator{$\;\;$}
\theoremsymbol{}
\renewtheorem{theorem}{Theorem}
\renewtheorem{proposition}[theorem]{Proposition}
\renewtheorem{lemma}[theorem]{Lemma}
\renewtheorem{corollary}[theorem]{Corollary}

\theoremstyle{plain}
\theorembodyfont{\upshape}
\renewtheorem{remark}{Remark}
\theoremsymbol{$\diamond$}
\renewtheorem{definition}{Definition}

\theoremstyle{nonumberplain}
\theoremheaderfont{\scshape}
\theorembodyfont{\upshape}
\theoremsymbol{$\square$}
\renewtheorem{proof}{Proof}

\newcommand{\N}{\mathbb{N}}
\newcommand{\R}{\mathbb{R}}
\newcommand{\C}{\mathbb{C}}
\newcommand{\cI}{\mathcal{I}}
\newcommand{\cM}{\mathcal{M}}
\newcommand{\cO}{\mathcal{O}}
\newcommand{\cU}{\mathcal{U}}
\newcommand{\cV}{\mathcal{V}}
\newcommand{\setlr}[1]{\left\{#1\right\}}
\newcommand{\set}[1]{\{#1\}}
\newcommand{\midvert}{\;\middle\vert\;}
\newcommand{\abs}[1]{\lvert#1\rvert}
\newcommand{\opint}[1]{\mathopen]{#1}\mathclose[}
\newcommand{\defl}{\vcentcolon=}
\newcommand{\defr}{=\vcentcolon}
\newcommand{\Spec}{\text{Spec}}
\newcommand\myphantom{\ensuremath{\vphantom{\left(1-\frac{S}{K_2}\right)}}}

\title{Tikhonov-Fenichel Reductions and their Application to a Novel Modelling Approach for Mutualism}
\author{Johannes Apelt \and Volkmar Liebscher}
\date{{\small Institute of Mathematics and Computer Science, University of
Greifswald,\\ Walther-Rathenau-Straße 47, 17489 Greifswald, Germany}
\\[\baselineskip] September 20, 2025}

\begin{document}

\maketitle

\begin{abstract}
  When formulating a model there is a trade-off between model complexity and
  (biological) realism. 
  In the present paper we demonstrate how model reduction from a precise
  mechanistic \enquote{super model} to simpler conceptual models using
  Tikhonov--Fenichel reductions, an algebraic approach to singular perturbation
  theory, can mitigate this problem.
  Compared to traditional methods for time scale separations (Tikhonov's theorem,
  quasi-steady state assumption), Tikhonov--Fenichel reductions have the
  advantage that we can compute a reduction directly for a separation of rates
  into slow and fast ones instead of a separation of components of the system. 
  Moreover, we can find all such reductions algorithmically.

  In this work we use Tikhonov--Fenichel reductions to analyse a
  mutualism model tailored towards lichens with an explicit description of the
  interaction. 
  We find: 
  (1)
  the implicit description of the interaction given in the reductions by
  interaction terms (functional responses) varies depending on the scenario,
  (2) 
  there is a tendency for the mycobiont, an obligate mutualist, to always
  benefit from the interaction while it can be detrimental for the photobiont,
  a facultative mutualist, depending on the parameters, 
  (3)
  our model is capable of describing the shift from mutualism to parasitism, 
  (4) 
  via numerical analyis, that our model experiences bistability with multiple
  stable fixed points in the interior of the first orthant.
  To analyse the reductions we formalize and discuss a mathematical criterion
  that categorizes two-species interactions. 
  Throughout the paper we focus on the relation between the mathematics behind
  Tikhonov--Fenichel reductions and their biological interpretation.
\end{abstract}

\section{Introduction}
\label{sec:intro}
Mutualism, an interspecific interaction that increases the fitness of both
partners involved, is present in virtually every ecosystem on earth
\cite{bronstein2015}. 
There exists barely any species that does not participate in a mutualistic
interaction of some form~\cite{bronstein2015,janzen1985}. 
Despite this fact, research focussed on individual examples of mutualism
separately instead of a general framework for a long time. 
In particular, the study of mutualism models lagged behind that of antagonistic
interactions \cite{bronstein1994a,bronstein2015,hale2021}. 

Early models for two populations interacting mutualistically, 
such as the model by~\textcite{gause1935}, 
are built upon the Lotka--Volterra framework with linear interaction terms
\cite{hale2021}.
These models are prone to show unbounded growth~\cite{hale2021} due to the
inherent positive feedback between mutualists, which May (1981) aptly coined an
\enquote{orgy of mutual benefaction}~\cite{murray2002}. 
It thus became apparent that realistic mutualism models need to be more
sophisticated. 
In particular, researchers investigated the mechanisms causing a saturation of
the positive feedback, which prevents population sizes from becoming infinite
\cite{hale2021}.
Such a saturation can be caused by intraspecific density dependence, i.e.\ a
decrease in births or an increase in deaths (or both) for one mutualist when
its population size increases, as for instance discussed
by~\textcite{wolin1984}. 
Another possibility used in various models is interspecific density dependence,
which in this case means the net benefit for a mutualist saturates or even
decreases with the abundance of its partner~\cite{hale2021}. 
An early example for this is the model by~\textcite{wright1989}, which uses a
Holling type II functional response~\cite{holling1959} as the interaction
term.
More recently, mechanistic models derived from considering the costs and
benefits of mutualism were proposed, e.g.\ by~\textcite{holland2002}. 
\textcite{holland2010} furthermore introduced the consumer-resource framework
for mutualism models, which considers the goods that are exchanged between
mutualists (uni- or bidirectional as well as indirect). 

Today, many mutualism models (for two populations) have been presented, most of
which rely on one or several concepts mentioned above. 
An overview is given by~\textcite{hale2021} (see also~\cite{georgescu2019}).
It is perhaps surprising that these models mostly yield consistent predictions,
considering that the model derivations and mechanisms are so different
\cite{hale2021}.  
However, justifying why a particular model is suitable for representing a real
world system constitutes a difficult task~\cite{getz2018}.
Even if the mechanisms at work can be identified, their abstraction as
mathematical formulae is somewhat arbitrary. 
Consider for instance the plethora of functional responses that have been
presented (see e.g.~\cite{jeschke2002} and references therein). 
How can we decide which one is particularly suited --- especially given that
many of them are qualitatively similar? 
In addition, (quantitative) predictive power is not necessarily a good measure
of how adequate a model is, since it depends on the quality of the available
data. 
Models may be useful as explorative tools and as such not
primarily meant to fit real data~\cite{odenbaugh2005,getz2018}.
From a mathematical perspective, one might want to require structural stability
(see e.g.~\cite{wiggins2003}) for a model to be a good model, but this
property alone is not a sufficient criterion.
We think the most important step towards formulating a sensible model is to
carefully examine the relevant biological mechanisms, which are then translated
directly into mathematics as explicitly as possible.
The disadvantage of this approach is that such models tend to be very complex
and difficult to analyse.

Here we present a novel approach that overcomes some of the difficulties of
modelling population dynamics --- in particular for mutualistic interactions. 
We closely follow the ideas presented by~\textcite{turchin2013}, namely that
general, well established assumptions (akin to axioms) should be used as
building blocks to derive models in theoretical ecology, e.g.\ by mathematical
reasoning.
Similarly,~\textcite{metz2005} argues that specific models should be embedded
\enquote{in a larger class of models, some members of which connect more
directly to the real biological world.}
These types of models will be called \emph{super models} throughout this paper.
The approach presented here yields a justification for using simpler conceptual
models depending on the particular setting via a super model.
This is also in line with ideas for model improvement proposed
by~\textcite{getz2018}.
In particular, we apply mathematically sound methods, which are capable of
semi-automatically performing what they call \enquote{coarse graining}. 

We derive our super model for mutualism from first principles by considering
relevant processes on the level of individuals.
This yields the actual system of ODEs as usual with the analogue of the
law of mass action~\cite{dilao2000,hutchinson2007,snyder2017,voit2015}. 
Since mutualisms are rather diverse concerning the underlying mechanisms that
constitute costs and benefits~\cite{bronstein2015,janzen1985}, we present our
ideas with one well studied example to facilitate interpretation --- namely the
lichen symbiosis. 
Note however, that the approach can be used to describe all types of mutualism
and that mathematical techniques in this paper are generally applicable to
polynomial (and to a lesser extent also rational) ODE systems --- and thus for
many models describing population dynamics. 
The essential idea for modelling the population dynamics of two mutualistic
species is to consider which individuals are actually involved in such an
interaction at a given point in time. 
Thus, individuals are either in autarkic or mutualistic state. 

The starting point for our modelling approach is the system
\begin{equation}
  \label{eq:lichenmodel}
  \begin{aligned}
    \dot{H} &= - \delta_1 H - \eta SH + \mu_1C\left(1-\frac{H}{K_1}\right) \\
    \dot{S} &= \beta_2 S \left(1-\frac{S}{K_2}\right) - \delta_2 S - \eta S H + \mu_2 C \left(1-\frac{S}{K_2}\right) \\
    \dot{C} &= \beta_3 C \left(1-\frac{C}{K_3}\right) - \delta_3 C + \eta S H
  \end{aligned}
\end{equation}
describing the population sizes of a host $H$ and its symbiont $S$ in their
autarkic state, i.e.\ free-living without their partner, 
and the mutualistic complex $C$ representing their interaction. 
The interaction can increase the fitness of both partners, because it
contributes additional births of individuals in autarkic and mutualistic state,
which renders the interaction potentially mutualistic.
In case of the lichen symbiosis, the interpretation of the complex is
straightforward: It represents the actual lichen association.  
In other scenarios, such as pollination mutualisms or zoochory,
the complex simply indicates the state of the interacting individuals (e.g.\ the
subpopulation of plants being visited by a pollinator or seed disperser at a
point in time).

For the lichen symbiosis, $H$ represents the mycobiont (fungus), an ecologically
obligate mutualist~\cite{honegger1998,palmqvist2000}, and $S$ its
facultative
partner, the photobiont (algae or cyanobacteria)
\cite{honegger1998,sanders2021}.
According to this characterization, the model includes autarkic reproduction for
the symbiont but not for the host.
Furthermore, we allow the mutualistic complex to reproduce, since this can occur
in lichens via thallus fragmentation~\cite{honegger1993}.

From this super model we consider reductions towards a two-dimensional system,
which is the common way to represent mutualism between two partners (compare
e.g.~\cite{hale2021}).  
This leads to mathematically simpler models, which still capture the essential
features of the full System~\eqref{eq:lichenmodel}. 
For this, we use the toolbox of algebraic Tikhonov--Fenichel reductions mainly
developed by Goeke and Walcher in a series of publications in the context of
enzyme kinetics
\cite{goeke2012,goeke2013,goeke2013a,goeke2014,goeke2015,goeke2017}.  
These reductions rely on the presence of differing time scales (slow and fast
processes) emanating from of a small parameter $\varepsilon > 0$. 
The general framework is as in Tikhonov's theorem~\cite{tikhonov1952} and 
related work by~\textcite{fenichel1979}. 
The former roughly states that, under some assumptions, a system of ODEs in the
form
\begin{equation}
  \label{eq:tikhonovnormal}
\begin{aligned}
    \dot{x} &= f(x,y) + \cO(\varepsilon), & x(0) &= x_0, & x\in D &\subseteq \R^s \\
    \varepsilon \dot{y} &= g(x,y) + \cO(\varepsilon), & y(0) &= y_0, & y\in G &\subseteq\R^r
\end{aligned}
\end{equation}
converges to
\begin{equation}
  \label{eq:tikhonovreduced}
\begin{aligned}
    \dot{x} &= f(x,y), & x(0) &= x_0 \\
    0 &= g(x,y), & y(0) &= y_0
\end{aligned}
\end{equation}
as $\varepsilon\to0$. 
Precise statements can be found in~\cite{verhulst2007}.
Due to this convergence, the essential behaviour for a particular separation of
rates remains the same as in the full system, i.e.\
the $s$-dimensional System~\eqref{eq:tikhonovreduced} is a good approximation for
the $(r+s)$-dimensional System~\eqref{eq:tikhonovnormal} if $\varepsilon$ is small. 
This approach has several advantages:
We can analyse the reduction instead of the full system, which is likely easier
because of the reduced dimension. 
Furthermore, the full system can potentially describe reality more accurately
due to an increased level of detail, which is conveyed to the reduction.
Note that this approach also naturally relates to the framework proposed by
\textcite{metz2005}, i.e.\ System~\eqref{eq:tikhonovnormal} being the super
model and System~\eqref{eq:tikhonovreduced} a particular scenario defined by
the slow--fast separation. 
The process of obtaining the reduction can be identified as coarse graining
as discussed by~\textcite{getz2018}.

In Tikhonov's theorem the system of ODEs is assumed to be in standard form
\eqref{eq:tikhonovnormal}, i.e.\ the choice of coordinates admits a slow--fast
separation of the state variables. 
It was already pointed out by~\textcite{fenichel1979} that this standard form
is not natural. 
A recent overview of singular perturbation theory in a coordinate-free setting
can be found in~\cite{wechselberger2020}.
Here, we use the results by Goeke and Walcher, which allows us to compute a
reduction independent of the choice of coordinates. 
But more importantly, since a model formulated following the law of mass action
is polynomial, this theory allows us to use tools from commutative algebra and
algebraic geometry to facilitate the computation of such reductions. 
In particular, we can determine all critical parameters admitting
a reduction as in \Cref{thm:reduction} for a given ODE system algorithmically
\cite{goeke2015}.

Using this algebraic approach, \textcite{kruff2019} showed that one can use
time scale reductions to derive the famous model by~\textcite{rosenzweig1963}
from first principles.
In contrast to the usual argument of a fixed balance in handling and searching
time of the predator \cite{holling1959}, 
this approach also accounts for dynamical properties of the system.
Similarly, \textcite{revilla2015} used an ad hoc time scale approach to derive
a two-species mutualism model from a model with an explicit description of
resources. 
However, as we will see, the algebraic approach to model reductions is much more
flexible and systematic.

The objective of the present paper is not so much to justify an existing model,
but to explore the space of models (i.e.\ reduced systems) that belong to the
class of our super model for mutualism.
In addition, System~\eqref{eq:lichenmodel} proved to be mathematically more
challenging than the predator--prey model in~\textcite{kruff2019}. 
Therefore, we will discuss the algorithmic toolbox of Tikhonov--Fenichel
reductions in more detail. 
To find and compute reductions in practice, we developed the \texttt{Julia}
\cite{bezanson2017} package \texttt{TikhonovFenichelReductions.jl}
\cite{apelt2024}, which includes subtle extensions to the work
by~\textcite{goeke2015} about algorithmic aspects of Tikhonov--Fenichel
reduction theory, which turned out to be particularly helpful in tackling this
problem.

With this work we want to demonstrate the strength of this approach in deriving
conceptual models.
Ultimately, we hope that the ideas presented here remind and help modellers to
seek a solid (mathematical) justification for the choice of a particular model
in various circumstances.

The paper is organized as follows. 
In \Cref{sec:mutualismmodel} we consider the derivation and some basic
properties of System~\eqref{eq:lichenmodel}. 
\Cref{sec:methods-reductions} contains a brief description of the underlying
mathematics used to derive reduced systems from \eqref{eq:lichenmodel}. 
Readers who are familiar with or not interested in the mathematics can skip
this section.
In  \Cref{sec:methods-analysis} we introduce a mathematical characterization of
mutualism and discuss concepts used to analyse the reductions of
System~\eqref{eq:lichenmodel}, of which some interesting cases are discussed in
\Cref{sec:reductions}. 
Appendix \ref{apx:reductions} contains an introduction to relevant concepts from
computational algebra and algebraic geometry, proofs and mathematical
derivations as well as supplementary material. 

\section{A Super Model for the Lichen Mutualism}
\label{sec:mutualismmodel}
For the present paper, we only consider a particular type of mutualism ---
namely that of the lichen symbiosis. 
However, the general modelling framework is also applicable to other types of
interactions. 
The main idea is to describe the interaction of individuals explicitly:
The population of each partner is partitioned into individuals that actually
participate in a mutualistic interaction and those who do not (at a specific
point in time).
We will call these \emph{mutualistic} and \emph{autarkic} individuals,
respectively.
Assuming that a mutualistic interaction always occurs between a fixed number of
individuals (or with a fixed ratio of each species' biomass), 
the population dynamics can be described with three compartments: 
The populations of individuals in autarkic state for both partner species and
the population of individuals in mutualistic state, called the mutualistic
complex. 
The model is then based on the following assumptions:
\begin{enumerate}[leftmargin=*, label=\roman*, ref=\roman*, widest=viii, align=left]
  \item \label{assumpt:interactingpops} Two populations are interacting
    mutualistically in one ecosystem with constant environmental conditions.
  \item \label{assumpt:conttime} Generations are overlapping and reproduction
    occurs homogeneously in time.
  \item \label{assumpt:densitydep} Populations of both species and their
    complex experience negative density dependence effects.
  \item \label{assumpt:migration} There is no migration.
  \item \label{assumpt:structure} The populations are well-mixed and there is no
    additional structure within them (such as age, sex or space related).
  \item \label{assumpt:resourcesdiffer} Both autarkic populations and the
    mutualistic complex are limited by different resources, i.e.\ there is no
    competition.
  \item \label{assumpt:benficialinteraction} Both partners benefit from being
    in a mutualistic relation via increased reproduction.
  \item \label{assumpt:1:1} There is a 1:1 relation between both partners
    within the mutualistic complex.
\end{enumerate}
These assumptions justify the use of ordinary differential equations (continuous
time) given that the number of individuals is sufficiently large to prevent
stochastic effects such as random extinctions and genetic drift to be important
drivers of the population dynamics. 
Note that in cases with imbalanced ratio within the mutualistic complex,
assumption~\ref{assumpt:1:1} can be satisfied by appropriate rescaling of
population sizes. 
Overall, these assumptions are intended to assure that the population dynamics
are predominantly affected by the potentially mutualistic interaction.
Since populations are not able to grow indefinitely~\cite{turchin2013},
we assume that births in all populations occur logistically.
Assumption~\ref{assumpt:structure} is obviously unrealistic, but commonly used
and keeps the model formulation feasible. 
The underlying idea is that differences in individuals are averaged out over the
population.
It might be difficult to judge whether assumption~\ref{assumpt:resourcesdiffer}
is satisfied in a real world system,
but in case of the lichen symbiosis we argue that this is a reasonable
simplification.
The mycobiont as a C-heterotrophic life form is mostly carbon limited
\cite{honegger1993}, whilst the lichen association is primarily limited by
water
availability and light~\cite{palmqvist2000}. 
The latter is generally also true for the photobiont, but we assume that lichens
and autarkic photobionts do not interfere, because they exist on different
spatial scales --- in fact, free-living photobionts may also occur in the direct
vicinity of lichens~\cite{peksa2022}.
Thus, we assume that there is no competition and the carrying capacities for all
populations are independent.

\begin{table}[t]
  \caption{Processes considered in the mutualism model for the lichen
    symbiosis. In the columns \enquote{Process} and \enquote{Reaction}, $H$,
    $S$ and $C$ denote one individual from the population instead of the
    population sizes.}
  \label{tab:processes}
  \begin{tabularx}{\textwidth}{@{}r X X X @{}}
    \toprule
    & Process & Reaction & Rate \\
    \midrule
    1 & $S$ gives birth & $S\rightarrow S + S$ 
      & $\beta_2\left(1-S/K_2\right)$ \\
    2 & $C$ gives birth & $C \rightarrow C + C$ 
      & $\beta_3\left(1-C/K_3\right)$ \\
    3 & $H$ gives birth from $C$ & $C \rightarrow C+H$ 
      & $\mu_1\left(1-H/K_1\right)$ \\ 
    4 & $S$ gives birth from $C$ & $C \rightarrow C+S$ 
      & $\mu_2\left(1-S/K_2\right)$ \\
    5 & $H$ dies & $H \rightarrow \emptyset$ 
      & $\delta_1$ \\
    6 & $S$ dies & $S \rightarrow \emptyset$ 
      & $\delta_2$ \\
    7 & $C$ dies & $C \rightarrow \emptyset$ 
      & $\delta_3$ \\
    8 & $C$ is formed & $H+S \rightarrow C$ 
      & $\eta$ \\
      \bottomrule
  \end{tabularx}
\end{table}
For a concrete model for the lichen symbiosis, we consider the processes given 
in \Cref{tab:processes}.
Note that we assume that mycobionts cannot reproduce autarkically,
because they are ecologically obligate mutualists --- although they are not
physiologically dependent on their
symbionts~\cite{honegger1998,palmqvist2000}.
Reproduction of the whole lichen complex can be found in nature for instance via
thallus fragmentation or symbiotic propagules~\cite{honegger1993} and is thus
reflected in the model.
The possibility for the symbiont to reproduce from within the lichen complex
(process 4 in \Cref{tab:processes}) is based on its ability to escape a lichen
thallus that has been damaged, e.g.\ due to heavy rain or predation
\cite{peksa2022}.

\begin{table}[h]
  \caption{Parameters of system~\protect\eqref{eq:lichenmodel} and their interpretation.}
  \label{tab:parameters}
  \begin{tabularx}{\textwidth}{@{} l X @{}}
    \toprule
    Parameter & Interpretation \\
    \midrule
    $\beta_i$ & Per-capita birth rate  \\
    $\delta_i$ & Per-capita death rate \\
    $K_i$ & capacity defined by resource availability \\
    $\eta$ & Rate of formation of mutualistic complex $C$ \\
    $\mu_i$ & Per-capita birth rate (from complex into autarkic state) \\
    \bottomrule
  \end{tabularx}
\end{table}
Following mass action kinetics, one can derive System~\eqref{eq:lichenmodel}
with the processes and their rates given in \Cref{tab:processes} as described
in~\cite[Eq.\ 2.1 and 2.2]{dilao2000}.
The parameters are explained in \Cref{tab:parameters}.
Note that the law of mass action requires constant rates, which is not the case
for the birth processes.
One can derive the logistic growth by considering the resources explicitly
\cite{dilao2000}, but we may also assume that births occur with constant rate
and additional deaths occur upon contact due to intraspecific competition. 
Then, instead of using the logistic growth rate directly, we can consider
process 1 as the combination of two processes:
\begin{equation*}
     S \xrightleftharpoons[\frac{\beta_2}{K_2}]{\beta_2} S + S 
\end{equation*}
(and analogously for the other birth processes), 
which allows us to apply the law of mass action directly and also yields
System~\eqref{eq:lichenmodel}.

The deterministic and time-continuous framework implies that the usual
assumption of sufficiently large and well-mixed populations is satisfied.
In case population sizes are small, one should describe the dynamics as a
stochastic process to depict the influence of random fluctuations (extreme
weather events, genetic drift, etc.) and allow for local extinction. 
However, this can also be used as an intermediate step to derive the ODE system
as a large volume limit from a Markov process with jumps according to
\Cref{tab:processes}~\cite{kurtz1980}.

\subsection{Some Properties of System~\protect\eqref{eq:lichenmodel}}
\label{sec:properties_lichenmodel}
We begin by summarizing some results for System~\eqref{eq:lichenmodel}. 
Proofs of the statements in this section can be found in
Appendix \ref{apx:proofsandcomputations}.
A first sanity check is to consider whether solutions can explode.
This is especially important for mutualism models due to their tendency towards
unlimited growth~\cite{murray2002}.
Here however, solutions are bounded.

\begin{figure}[t]
  \begin{center}
    \pgfplotstableread[col sep=comma]{simulation_f_bistability.csv}\tabbistability
    \begin{tikzpicture}
      \begin{axis}[
        width=0.75\textwidth, height=0.5\textwidth,
        xlabel=$t$,
        legend entries={$H$,$S$,$C$},
        legend cell align={left},
        legend style={
          at={(1.04,1)},
          anchor=north west,
          draw=none
        },
        grid=major,
        axis lines=left,
        xmin=0, ymin=0,
        enlargelimits=true,
        axis background/.style={fill=gray!05}
        ]
        \addplot [orange!90!brown, semithick] table [x=t,y=Ha] {\tabbistability};
        \addplot [teal!70!green, semithick] table [x=t,y=Sa] {\tabbistability};
        \addplot [cyan!40!blue, semithick] table [x=t,y=Ca] {\tabbistability};
        \addplot [orange!90!brown, thick, dashed] table [x=t,y=Hb] {\tabbistability};
        \addplot [teal!70!green, thick, dashed] table [x=t,y=Sb] {\tabbistability};
        \addplot [cyan!40!blue, thick, dashed] table [x=t,y=Cb] {\tabbistability};
      \end{axis}
    \end{tikzpicture}
  \end{center}
  \caption{Simulated time series for System~\eqref{eq:lichenmodel} in a
    bistable scenario with parameters $\beta_2 = 9$, $\beta_3 = 6$, $\delta_1
    = 4$, $\delta_2 = 2$, $\delta_3 = 7$, $\mu_1 = 5$, $\mu_2 = 2$, $\eta =
    20$, $K_1 = K_2 = 10$ and $ K_3 = 6$.
    The initial values are $H(0)=S(0)=0.1$ for both simulations, but
    $C(0)=0.1$ for the one shown in solid lines and $C(0)=4$ for the one with
    dashed lines.
    Note that the parameters satisfy condition \ref{thm:existenceinteriorfp3}
    in \Cref{thm:existenceinteriorfp}.
  \label{fig:lichenmodelbistable}}
\end{figure}

\begin{theorem}
\label{thm:solutionbounded}
  Consider System~\eqref{eq:lichenmodel} and let $D=[0, K_1] \times [0, K_2]
  \times [0, \tilde{K}_3]$ with 
  \begin{equation*}
    \tilde{K}_3 \defl \frac{1}{2} \left( \frac{\beta_3 - \delta_3}{\beta_3} K_3 
    + \sqrt{\left(\frac{\beta_3-\delta_3}{\beta_3}\right)^2 {K_3}^2 + 8\frac{\eta}{\beta_3} K_1 K_2K_3} \right) .
\end{equation*} 
  Then $D$ is positively invariant under \eqref{eq:lichenmodel}.
\end{theorem}

Computing all fixed points of System~\eqref{eq:lichenmodel} is not
straightforward, since those in the interior of $D$ as defined in
\Cref{thm:solutionbounded} are the roots of a fourth-order polynomial, 
where the coefficients are rational expressions of the parameters. 
However, in some cases we can guarantee the existence of an interior fixed
point.
\begin{theorem}
  \label{thm:existenceinteriorfp}
  The only fixed points in $\R_{\geq 0}^3$ with some components vanishing are
  $(0,0,0)$ and $\left(0, \frac{\beta_2-\delta_2}{\beta_2} K_2, 0 \right)$.
  Any fixed point $(H^\star, S^\star, C^\star)$ in the interior of $D$ must
  satisfy 
  \begin{equation*} 
    \frac{\beta_3 - \delta_3}{\beta_3} K_3 < C^\star < \frac{\beta_3 - \delta_3 + \mu_1}{\beta_3} K_3 
  \end{equation*}
  In each of the following cases there exists at least one interior fixed
  point.
  \begin{enumerate}[label=(\roman*), widest=(iii)]
    \item \label{thm:existenceinteriorfp1} 
      $\beta_3 > \delta_3$
    \item \label{thm:existenceinteriorfp2}
      $\beta_3 = \delta_3$ and $\beta_2-\delta_2 > -\frac{\eta \mu_1 \mu_2 K_3}{\beta_3 \delta_1}$
    \item \label{thm:existenceinteriorfp3}
      $-\mu_1 < \beta_3 - \delta_3 < 0$ and $\beta_2 - \delta_2 > -\frac{\beta_2\delta_1(\beta_3-\delta_3)}{\eta K_2 (\beta_3-\delta_3+\mu_1)}$
  \end{enumerate}
\end{theorem}

Interestingly, the interior fixed point is not necessarily unique. 
There are scenarios in which System~\eqref{eq:lichenmodel} shows bistability.
More precisely, there exist three interior fixed points of which two are stable. 
The numerical simulations depicted in  \Cref{fig:lichenmodelbistable} demonstrate
this behaviour of the system and we will see later that bistability is
preserved in some reductions (compare \Cref{fig:red12_bistability}).

This behaviour is -- to the authors' knowledge -- not very common in
population dynamics, but known to occur in chemical
systems~\cite{craciun2005}. 
Typically, when bistability is observed in mutualism models, one of the fixed
points is characterized by the vanishing of one population (see e.g.\
\cite{vet2018} and  \Cref{fig:reduction21}).
The possibility for multiple interior fixed points in mutualism models for two
populations has been noted by~\textcite{brauer1985}, but no concrete example
showing
this behaviour is given.
\textcite{thompson2006} demonstrated that this phenomenon can occur in models with
immigration.

\section{Tikhonov-Fenichel Reduction Theory}
\label{sec:methods-reductions}
In this section we will briefly consider Tikhonov's
theorem \cite{tikhonov1952} and discuss the algebraic extension building on
Fenichel's work~\cite{fenichel1979}, which yield a convenient way to find and
compute model reductions.
A short overview of the essential concepts from commutative algebra and
algebraic geometry needed here is given in Appendix
\ref{apx:mathematicalpreliminaries}.
Further details can be found in~\cite{kunz2013} and~\cite{cox2015}.
Note that some notions in algebraic geometry are used with subtle but
important differences --- mainly due to problems arising when working over a
field that is not algebraically closed and because of different naming
conventions. 
Here we mostly follow~\textcite{kunz2013}.
Finally, we will discuss how this theory can be applied in practice. 

\subsection{Tikhonov's Theorem and the Work of Fenichel}
The general framework of Tikhonov--Fenichel reductions is singular perturbation
theory. 
More precisely, we consider Tikhonov's theorem as in~\cite{verhulst2007}.
Roughly speaking, it states that (under certain assumptions), for
$\varepsilon>0$, solutions of a system in the form \eqref{eq:tikhonovnormal}
converge to solutions for the corresponding System~\eqref{eq:tikhonovreduced} as
$\varepsilon \to 0$ on some (possibly finite) time interval. 
Note that Tikhonov's theorem is even more general and not restricted to the
autonomous case. 
However, for us this restriction is not problematic --- in fact, it allows us to
use the theory developed by~\textcite{fenichel1979} in a straightforward manner,
which yields a geometric interpretation and shall prove to be quite useful.

We closely follow the terminology in~\cite{goeke2014} and say that a system
given as in \eqref{eq:tikhonovnormal} is in \emph{Tikhonov normal form}.
System~\eqref{eq:tikhonovreduced} is called a \emph{Tikhonov--Fenichel reduction}
of the full System~\eqref{eq:tikhonovnormal} (\emph{reduced system} or
\emph{reduction} in short). 
We can interpret the small parameter $\varepsilon$ as the conversion rate
between the characteristic time scales that $x$, the slow variable, and $y$, the
fast variable, evolve on. 
Note that System~\eqref{eq:tikhonovreduced} is restricted to the set $\cM_0
\defl \set{(x,y) \mid g(x,y)=0}$, the so-called \emph{slow manifold}
(sometimes also \emph{critical manifold}), which is an invariant set of
System~\eqref{eq:tikhonovreduced}. 
This means a subset of all components is sufficient to describe the flow of the
full system reasonably well (on the time interval for which convergence is
guaranteed). 
More precisely, one of Fenichel's theorems states that if $\cM_0$ is
attractive, solutions of
System~\eqref{eq:tikhonovnormal} will stay on a locally invariant manifold
$\cM_{\varepsilon}$ within $\cO(\varepsilon)$ of
$\cM_0$, which is diffeomorphic to $\cM_0$, for $\varepsilon>0$
sufficiently small~\cite{hek2010}.
Thus, the essential behaviour of the full system is already captured by
$\cM_0$ and System~\eqref{eq:tikhonovreduced}.
This explains why such a reduced system is useful in practice: It shows
essentially the same dynamic behaviour as the full system and is potentially
much simpler to handle mathematically due to its lower dimension.
For precise statements and a thorough description of the theory we refer the
interested reader to~\cite{verhulst2007} and~\cite{hek2010} (see also
\cite{wechselberger2020}).

Note that the time interval on which the reduced system is convergent is not
known a priori and may be finite. 
Sometimes this means that the reduced system shows a qualitatively different
long time behaviour due to its approximative nature.
In this case, one may also compute higher order approximations of the slow
manifold and corresponding reduction, as for instance demonstrated
by~\textcite{poggiale2004}.

Although quite powerful, Tikhonov's theorem can only be applied if a system of
ODEs is already in Tikhonov normal form. 
This means the components must be divided into slow and fast ones, which
is often called quasi-steady state assumption.
Such a separation might be obtained with prior knowledge, but in general this
will not be clear. 
For ODE systems with polynomial (and to a lesser extent rational or analytic)
RHS, one solution to this problem is to apply the algebraic reduction theory
presented
in~\cite{goeke2012,goeke2013,goeke2013a,goeke2014} and \cite{goeke2015,goeke2017} (see
also
\cite{wechselberger2020}).
This enables us to obtain all reductions that arise from a slow--fast separation
of processes (instead of the separation of components as in Tikhonov's theorem). 
Essentially, a Tikhonov--Fenichel reduction is a reduction in the sense of
Tikhonov for a system given in the form
\begin{equation}
  \label{eq:generalsystem}
  \dot{x} = f(x,\pi,\varepsilon) = f^{(0)}(x) + \varepsilon f^{(1)}(x) +
  \cO(\varepsilon^2), \quad x\in U, \pi\in\Pi ,
\end{equation}
where $U \subseteq \R^n$ and $\Pi \subseteq \R^m$ are open and $f$ is
analytic.
Note that $f^{(0)}$ and $f^{(1)}$ still depend on the base parameters $\pi$,
but we interpret them as fixed, which will make the notation more elegant.

The basic question is whether we can find a transformation into Tikhonov normal
form such that we can apply Tikhonov's theorem. 
It turns out, that we do not need to find such a transformation explicitly, as
we will see in the next section.  

\subsection{Tikhonov--Fenichel Reductions}
The following main result from~\cite{goeke2014} allows us to compute the
reduced
system in Tikhonov's theorem in a straightforward manner. 
Note that we will later assume that $f$ in~\eqref{eq:generalsystem} is
polynomial in order to find reductions systematically, but \Cref{thm:reduction}
also works if $f$ is analytic~\cite{goeke2014}.
$\cV(F)$ denotes the affine variety of a set or vector of polynomials $F$, which
in this context is the set of common roots of the polynomials in $F$.
Note that the existence of a nonsingular point in an affine variety implies
that it is locally a manifold.
Further details can be found in Appendix~\ref{apx:mathematicalpreliminaries}.

\begin{theorem}[Tikhonov--Fenichel Reduction{~\cite[][Thm.\ 1]{goeke2014}}]
 \label{thm:reduction}
  Let $f$ in System~\eqref{eq:generalsystem} be rational and $x_0 \in \R^n$ a
  nonsingular point in $\cV(f^{(0)})$. 
  Assume furthermore that there exists the direct sum decomposition
\begin{equation}
    \label{eq:directsum}
    \R^n = \textup{Ker}\; Df^{(0)}(x_0) \oplus \textup{Im}\; Df^{(0)}(x_0). 
\end{equation}
 Then the following hold.
\begin{enumerate}[label=(\roman*), widest=(ii)]
  \item \label{thm:reduction-composition} 
      Let $r\defl \textup{rank} \; Df^{(0)}(x_0)$ and $s\defl n-r$. 
      There exists a Zariski neighbourhood $\cU_{x_0}$ of $x_0$ and
      matrices $\psi(x)\in\R(x)^{r\times1}$ and $P(x)\in\R(x)^{n\times r}$
      admitting the product decomposition
\begin{equation}
   \label{eq:P_psi_composition}
   f^{(0)}(x) = P(x)\psi(x), \quad x\in\cU_{x_0},
\end{equation}
  with $\textup{rank}\; P(x_0) = \textup{rank}\; D\psi(x_0) = r$ and
\begin{equation}
   \label{eq:f_psi_vanishingsets}
   \cV(f^{(0)}) \cap \cU_{x_0} = \cV(\psi) \cap \cU_{x_0},
\end{equation}
  which is an $s$-dimensional submanifold.
  \item \label{thm:reduction-projection} 
      There exists a formal Tikhonov--Fenichel reduction onto an $s$-dimen\-sional
      Zariski neighbourhood $\tilde{\cU}_{x_0}\subseteq
      \cV(f^{(0)})$ of $x_0$ given by
\begin{equation}
   \label{eq:reduction}
      \dot{x} = \left[1_n - P(x)A(x)^{-1} D\psi(x)\right] f^{(1)}(x), 
      \quad x\in\tilde{\cU}_{x_0},
\end{equation}
  with the invertible matrix
  \begin{equation*} A(x) \defl D\psi(x)P(x) \in\R(x)^{r\times r}
\end{equation*}
  and $n=r+s$.
    Furthermore, $\tilde{\cU}_{x_0}$ is an invariant set under \eqref{eq:reduction}. 
  \item \label{thm:reduction3} 
      If all nonzero eigenvalues of $Df^{(0)}(x_0)$ have negative real part,
      System~\eqref{eq:reduction} restricted to the slow manifold \,
      $\tilde{\cU}_{x_0}$ corresponds to the reduction, i.e.\
      System~\eqref{eq:tikhonovreduced}, in Tikhonov's theorem.
\end{enumerate}
\end{theorem}
Therefore, to compute an $s$-dimensional reduction for System~\eqref{eq:generalsystem},
there has to exist an $s$-dimensional irreducible
component in $\cV(f^{(0)})$ containing a real nonsingular point $x_0$
admitting a direct sum decomposition of $\R^n$ into kernel and image of the
Jacobian of $f^{(0)}$ at $x_0$. 
Then, all that remains is to find a product decomposition for $f^{(0)}$
satisfying \eqref{eq:P_psi_composition} and \eqref{eq:f_psi_vanishingsets}. 
With that, the formal reduction is directly given by \eqref{eq:reduction} and we
can decide whether the slow manifold $\tilde{\cU}_{x_0}$ is attractive depending
on the eigenvalues of $Df^{(0)}(x_0)$.
If it is attractive, System~\eqref{eq:generalsystem} converges to
\eqref{eq:reduction} on some time interval as $\varepsilon\to0$ for initial
values sufficiently close to the slow manifold. 

\subsection{Finding Tikhonov--Fenichel Parameter Values}
\label{sec:findingtfpv}
With \Cref{thm:reduction} we can compute a formal reduction for an ODE system
that is already separated into slow and fast part as in~\eqref{eq:generalsystem}.
However, it is not always clear what a sensible slow--fast separation is a
priori. 
The major advantage of this approach to time scale separation is that it allows
us to use necessary conditions on the parameters to find all possible
reductions with methods from computational algebra~\cite{goeke2015}. 
This is the reason why we require $f$ to be polynomial, which we will assume in
the following. 
We begin with the notion of a critical parameter in our setting.
\begin{definition}[TFPV]
  A parameter $\pi'\in\Pi$ satisfying the conditions in \Cref{thm:reduction} for
  a system as in~\eqref{eq:generalsystem} is called a
  \emph{Tikhonov--Fenichel Parameter Value (TFPV)} for dimension $s$.
\end{definition}
Thus, TFPVs for dimension $s$ are precisely the parameter values that admit a
formal reduction to an $s$-dimensional slow manifold.
Note that the existence of the formal reduction given by~\eqref{eq:reduction} does not require attractivity of the slow
manifold as in condition \ref{thm:reduction3} in \Cref{thm:reduction}.  
However, attractivity is desired in practice, because it admits convergence as
in Tikhonov's theorem. 
Thus, a formal reduction is only a meaningful approximation of the original
system if the slow manifold is attractive.

Here, we introduce the search for TFPVs independent of attractivity, as we
consider asymptotic properties of the formal reduction once we chose a
particular point $x_0$ as in \Cref{thm:reduction}. 
Moreover, for a given polynomial ODE system, we may obtain TFPV candidates
admitting multiple reduced systems, as we will see later.
To find such critical parameter values, we can use the following
Proposition.

\begin{proposition}[Necessary Conditions for TFPVs~\cite{goeke2013,goeke2015}]
  \label{thm:tfpvnecessary}
  Consider System~\eqref{eq:generalsystem}.
  Let $\pi'\in\Pi$ be a TFPV for dimension $s$, $r\defl n-s$ and
  \begin{equation}
    \label{eq:charpoly}
    \chi_{x,\pi}(\tau) = \tau^n + \sigma_{n-1}(x,\pi) \tau^{n-1} +\cdots+ \sigma_1(x,\pi) \tau + \sigma_0(x,\pi)
  \end{equation}
  the characteristic polynomial of $D_1 f(x,\pi)$. 
  Then there exists $x_0\in U$ with the following properties:
  \begin{enumerate}[label=(\roman*), widest=(iii)]
    \item \label{thm:tfpvnecessary1} $f(x_0,\pi') = 0$
    \item \label{thm:tfpvnecessary2} For $k>r$ the determinant of each $k\times
      k$ minor of $D_1 f(x_0,\pi')$ vanishes.
    \item \label{thm:tfpvnecessary3} $\sigma_s(x_0,\pi') \neq 0$
  \end{enumerate}
\end{proposition}
If we know the dimension of the affine variety $\cV(f(\cdot,\pi'))$ (taken as a
subset of $\R^n$) for a TFPV candidate $\pi'$, we can guarantee the
existence of a formal reduction as follows.
\begin{proposition}[Sufficient Conditions for TFPVs~\cite{goeke2013}]
  \label{thm:tfpvsufficient}
  Consider System~\eqref{eq:generalsystem} and fix $\pi'\in\Pi$.
  Let $Y$ be an $s$-dimensional irreducible component of
  $\cV(f(\cdot,\pi'))$ and $x_0\in Y$.
  If \ref{thm:tfpvnecessary3} in \Cref{thm:tfpvnecessary} is satisfied, then
  $\pi'$ is a TFPV for dimension $s$.
\end{proposition}
We will discuss two different approaches for finding TFPVs for a given dimension
$s$ in the following.

\subsubsection{Exhaustive Search Using a Gröbner Basis Approach}
\label{sec:findingtfpvgroebner}
The first method relies on the computation of a Gröbner basis for an elimination
ordering.
Due to conditions \ref{thm:tfpvnecessary1} and \ref{thm:tfpvnecessary2} in
\Cref{thm:tfpvnecessary}, we are interested in the vanishing of $f$ and the
determinants of certain minors of $D_1 f$. 
Let $I \subseteq \R[x,\pi]$ be the ideal generated by these polynomials.
We search for TFPV candidates independently from the point $x_0$, because there
may exist multiple reductions onto different manifolds for the same TFPV.
We can therefore eliminate the variables $x$ from $I$, which can be achieved by
computing a Gröbner basis of $I$ with an elimination ordering for
$x$ (\cite{cox2015}, chapter 3).
Let $G$ be such a Gröbner basis. 
Then, the elimination ideal $I_{\pi} \defl I \cap \R[\pi]$ is generated by
the set of all polynomials in $G$ not containing any of the variables
$x_1,\dots,x_n$ (i.e.\ the state variables of the ODE system).
Note that the vanishing of the elimination ideal is necessary for the
satisfiability of the conditions in \Cref{thm:tfpvnecessary}, because one always
has $\cV(I) \subseteq \cV(I_{\pi})$.
Thus, every TFPV is contained in $\cV(I_{\pi})$.

This approach yields all possible TFPVs for a given ODE system with polynomial
RHS. 
This includes expressions of the parameters whose vanishing imply the vanishing
of $I_\pi$.
The disadvantage of this approach is that computing a Gröbner basis can be a
very costly task and we only get an implicit description of the set of TFPV
candidates. 
For high dimensional systems or a large number of parameters, this computation
may not even be feasible, especially if the reduction in dimension is large.
In such cases, we may still be able to compute all TFPVs of a special type with
the following method.

\subsubsection{Slow-Fast Separation of Rates}
\label{sec:findingsfsr}
The second procedure for finding TFPVs utilizes the fact that we can compute the
Krull dimension of the ideal $\langle f^{(0)} \rangle$
algorithmically, which is equal to the dimension of
$\cV_\C(f^{(0)})$.
Since the existence of a real nonsingular point in this variety is required in
\Cref{thm:reduction}, the Krull dimension also equals the real dimension in all
relevant cases.

In addition, we can obtain the irreducible components of an affine variety by
computing a minimal primary decomposition of its generating ideal.
This allows us to consider local properties of the variety and yields all the
potential slow manifolds (as affine varieties) on which a reduction may exist.
Using both facts, it becomes possible to find all TFPVs of a specific type
algorithmically --- namely \emph{slow--fast separations of rates}, which we define
as follows. 
\begin{definition}[Slow-fast Separation of Rates]
  \label{def:slow-fast-separation}
  Let $\pi\in\Pi\cap \R_{>0}^m$ be fix.
     A \emph{slow--fast separation of rates} $\tilde\pi=\tilde\pi(\pi,\varepsilon)$
  with base parameters $\pi_i$ is defined by 
  \begin{equation*}
    \tilde\pi_i \defl 
\begin{cases}
      \varepsilon \pi_i & i \in S \\ \pi_i & i \not\in S
\end{cases}
\end{equation*}
  for a nonempty index set $S\subset\set{1,\dots,m}$. 
  This means $(\pi_i)_{i\in S}$ are the small parameters corresponding to slow
  processes. 
  We will always write $\pi^\star\defl\tilde\pi(\pi,0)$.
\end{definition}
Slow--fast separations of rates are the TFPVs that we are typically most
interested in, since they directly relate to a time scale separation of
processes.
If the elimination ideal $I_\pi$ as above is a monomial ideal, every TFPV is a
slow--fast separation of rates.
Depending on the complexity of the input system and the drop in dimension, this
may not be very likely to occur in practice, but can be seen in e.g.\
\cite{kruff2019}.

To find reductions onto $s$-dimensional slow manifolds in practice, we can loop
over all $2^m-2$ possible slow--fast separations of rates and filter out TFPVs
using a refinement of the necessary conditions in \Cref{thm:tfpvnecessary}.
For this, we consider the components of $f$ as elements in $\R[x]$ (to work
symbolically with the parameters in a computer algebra system, we actually use
the polynomial ring $\R(\pi)[x]$). 
Let $\pi^\star$ be a slow--fast separation of rates and note that
$f^{(0)}=f(\cdot,\pi^\star)$ as in Eq.~\eqref{eq:generalsystem}.
The slow manifold is contained in a single irreducible component of
$\cV(f^{(0)})$ with dimension $s$.
Consider the minimal primary decomposition $(Q_i)_{i=1,\dots,k}$ of $\langle
f^{(0)} \rangle$, i.e.\  $\langle f^{(0)} \rangle = \bigcap_{i=1}^k Q_i$.
Then, each $\cV(Q_i)$ corresponds to an irreducible component of $\cV(f^{(0)})$
and we require that there exists $l$ such that $\dim Q_l = s$, since this is
equivalent to $\dim \cV_{\C}(Q_l) = s$ as desired.

The Jacobian $D_1 f(x,\pi^\star)$ can be computed symbolically.
However, we need a symbolic description of $D_1f(x',\pi^\star)$ for a
point $x'\in\cV_\C(Q_l)$ for which the characteristic polynomial can be
evaluated.
To assure condition \ref{thm:tfpvnecessary3} in \Cref{thm:tfpvnecessary} can be
satisfied, we can compute the normal form (denoted $\text{NF}$) of each entry of $D_1
f(x,\pi^\star)$ with respect to $\sqrt{Q_l}$ and then compute the characteristic
polynomial.
Note that it is important to take the radical of $Q_l$, because a polynomial may
be divisible by $\sqrt{Q_l}$ but not $Q_l$.
More precisely, Hilbert's Nullstellensatz tells us that
$\cI(\cV_\C(Q_l)) = \sqrt{Q_l}$, which implies that an entry
$p\in\R[x]$ of the Jacobian (or some of its terms) vanishes on
$\cV_\C(Q_l)$ exactly if it lies in $\sqrt{Q_l}$.
To see why it suffices to use normal forms, let $G=\set{g_1,\dots,g_k}$ be a
Gröbner Basis for $\sqrt{Q_l}$.
Then there exists a unique $r\in\R[x]$ and $q_1,\dots,q_k\in\R[x]$
such that $p = q_1g_1 +\cdots+ q_kg_k + r$. 
Thus, $p=r$ on $\cV_\C(Q_l)$.
This allows us to check whether $\sigma_s(x_0,\pi^\star)\neq 0$ can be satisfied
for any point $x_0\in \cV_\C(Q_l)$.

In summary, any slow--fast separation of rates $\pi^\star$ that is a TFPV
satisfies the following algorithmically accessible conditions:
\begin{enumerate}[label=(\roman*), widest=(iii)]
  \item $\dim \langle f(\cdot,\pi^\star) \rangle \geq s$
  \item Let $(Q_i)_{i=1,\dots,k}$ be a minimal primary decomposition for
    $\langle f(\cdot,\pi^\star) \rangle$, then $\exists
    l\in\set{1,\dots,k}:\dim Q_l=s$.
  \item The characteristic polynomial of
    \begin{equation*}
      \left(\text{NF}\left(\frac{\partial
          f_i}{\partial x_j}(x,\pi^\star),
      \sqrt{Q_l}\right)\right)_{(i,j)\in\set{1,\dots,n}^2}
\end{equation*}
    written as in
    Eq.~\eqref{eq:charpoly} satisfies $\sigma_s\neq0$.
\end{enumerate}

In order to satisfy the conditions in \Cref{thm:tfpvsufficient} we have to check
whether there exists a nonsingular point in $\cV_\R(Q_l)$, which needs to be
done manually.
However, we can postpone this step to the computation of the reduced systems,
because for this we usually want the slow manifold to be given explicitly as a
subset of $\R^n$, i.e.\ in parameterized form.
Note that this is not required to compute the formal reduction, but needed in
order to substitute variables according to the slow manifold
in~\eqref{eq:reduction} to obtain the system in local parameters. 
With this, one can easily check whether $D_1 f(x',\pi^\star)$ satisfies
condition \ref{thm:tfpvnecessary3} in \Cref{thm:tfpvnecessary} for a point
$x'\in\cV_{\R}(Q_l)$.
If that is the case, then the conditions in \Cref{thm:tfpvsufficient} are
satisfied. 

\subsubsection{Practical Considerations}
It turns out that the brute force approach for finding all slow--fast separations
of rates that are TFPVs is computationally less demanding than the approach
based on computing the elimination ideal via a Gröbner basis in most cases
(depending on the complexity of the input system and the drop in dimension).
If there exists a TFPV, which is not a slow--fast separation of rates but defined
by expressions in the parameters whose vanishing imply the vanishing of $I_\pi$,
we can introduce a new parameter for each of these expressions. 
Then, we rewrite the system of ODEs such that these new parameters become
slow--fast separations of rates for the new system.
Therefore, in most cases it is sufficient to deal with the computation of the
reductions for slow--fast separations.

To actually find TFPV candidates in practice, we developed the free and open
source software package
\texttt{TikhonovFenichelReductions.jl}~\cite{apelt2024},
which contains an implementation of both procedures discussed, i.e.\ for finding
all TFPVs and all slow--fast separations of rates that are TFPVs. 
Additionally, the package contains functions for computing the corresponding
reduced systems. 
The only manual steps required to compute a (formal) reduction are finding an
explicit description of the irreducible components of the affine variety
$\cV(f(\cdot,\pi^\star))$ as manifolds and a product decomposition as in
Eq.~\eqref{eq:P_psi_composition}. 
However, the package contains heuristics that can automate both tasks in many
cases.

\texttt{TikhonovFenichelReductions.jl} is written in \texttt{Julia}
\cite{bezanson2017} and mainly utilizes the package \texttt{Oscar.jl}
\cite{oscar2022,decker2025} (in particular one of its cornerstones
\texttt{Singular}~\cite{singular2022}).

\section{Analysis of Population Dynamics}
\label{sec:methods-analysis}
This section deals with the analysis of mathematical
models in terms of ecological properties. 
In particular, we introduce a mathematical criterion that allows us to
characterize when a given two-dimensional model describes mutualism. 
Additionally, we discuss some basic approaches needed to analyse reductions for
System~\eqref{eq:lichenmodel}.

\subsection{Mutualism Criteria}
Since a slow--fast separation of rates lies in an extreme region of the parameter
space, it is worthwhile to check whether the reduced system still describes
mutualism. 
We will see that this is indeed not always the case here.
In order to analyse the reductions in the following, we need to define what a
mutualistic model is mathematically.
For this we consider the general ODE system
\begin{equation}
  \label{eq:twospeciessystem}
  \dot{x} = f(x,y), \quad \dot{y} = g(x,y)
\end{equation}
for two populations of different species interacting. 
Let this system be defined on $D\subseteq\R^2$ and $f$ and $g$ continuous in
$\overline{D}$. 

\newpage
\begin{definition}[Strong Mutualism Criterion]
  \label{def:strongmutualismcriterion}
  If $f$ and $g$ are differentiable in $D$, we say that
  System~\eqref{eq:twospeciessystem} is \emph{strongly mutualistic} in $D$ if
  $\forall (x,y) \in D: $
\begin{align*}
    \frac{\partial f}{\partial y}(x,y) &\geq 0, \quad \frac{\partial f}{\partial y}\bigg\vert _D \not\equiv 0 \\
    \frac{\partial g}{\partial x}(x,y) &\geq 0, \quad \frac{\partial g}{\partial x}\bigg\vert _D \not\equiv 0
\end{align*}
\end{definition}
This definition is commonly used to characterize mutualism in mathematical
models (see e.g.~\cite{brauer1985,neuhauser2004,wang2011}).
For two-dimensional ODE systems this also aligns with the definition of a
cooperative system~\cite{smith1995}. 
However, since it relies on partial derivatives, it reflects the trend of the
effect of the interaction and not its magnitude.
We will therefore introduce and use another definition.
\begin{definition}[Mutualism Criterion]
  \label{def:mutualismcriterion}
  For a point $(x,y)\in D$ we define
  \[
    \underline{x}^D(y) \defl \inf\set{u\in\R \mid (u,y) \in D}
    \;\text{ and }\;
    \underline{y}^D(x) \defl \inf\set{v\in\R \mid (x,v) \in D}.
  \]
  We say that system \eqref{eq:twospeciessystem} is
  \emph{mutualistic} in $D$ if $\forall (x,y) \in D: $
  \begin{align*}
    b_1(x,y) \defl f(x,y) - f(x, \underline{y}^D(x)) &\geq 0, \quad b_1\big\vert_D \not \equiv 0 \\
    b_2(x,y) \defl g(x,y) - g(\underline{x}^D(y), y) &\geq 0, \quad b_2\big\vert_D \not \equiv 0
  \end{align*}
  and we call $b_1$ and $b_2$ the \emph{(net) benefit functions} for species
  $x$ and $y$, respectively.
\end{definition}
Note that a strongly mutualistic system is also mutualistic. 
A situation in which the benefit functions (or the partial derivatives) are zero
on $D$ is known as neutralism, and we speak of commensalism if only one of them
is identically zero. 
Other types of interactions can also be defined according to the sign of the
benefit functions.

\begin{figure}[t]
  \begin{center}
    \begin{tikzpicture}
      \draw[-Stealth] (-0.1,0) -- (7,0);
      \draw[-Stealth] (0,-0.1) -- (0,5);
      \draw (7,0) node[anchor=north] {$x$};
      \draw (0,5) node[anchor=east] {$y$};
      \draw[draw=teal, fill=teal, fill opacity=0.2] (0.5, 0.5) -- (6.5,4.5) -- (0.5, 4.5) -- cycle;
      \draw (5.8,4.8) node {$D$};
      \draw[thick, dotted, gray] (0,1.5) -- (7,1.5);
      \draw[thick, dotted, gray] (2,0) -- (2,5);
      \draw[thick, dotted, gray] (0,4) -- (7,4);
      \draw[thick, dotted, gray] (0.5,0) -- (0.5,5);
      \draw[thick, red] (2,1.5) -- (2,4) -- (0.5,4);
      \draw[fill=black] (2,4) circle[radius=1pt];
      \draw[fill=black] (2,1.5) circle[radius=1pt];
      \draw[fill=black] (0.5,4) circle[radius=1pt];
      \draw (2,0.1) -- (2,-0.1) node[anchor=north] {$\tilde{x}$};
      \draw (0.1,4) -- (-0.1,4) node[anchor=east] {$\tilde{y}$};
      \draw (0.5,0.1) -- (0.5,-0.1) node[anchor=north] {$\underline{x}^D(\tilde{y})$};
      \draw (0.1,1.5) -- (-0.1,1.5) node[anchor=east] {$\underline{y}^D(\tilde{x})$};
    \end{tikzpicture}
  \end{center}
  \caption{
    Schematic illustration of the mutualism criterion in
    \Cref{def:mutualismcriterion} for a triangular domain $D$. 
    The benefit function $b_1$ evaluates the difference in values that $f$
    attains along the vertical cut of $D$ through the point 
    $(\tilde{x},\tilde{y})$ and analogously for $b_2$ (indicated by the red
    lines).
  }
  \label{fig:mutualismcriterion}
\end{figure}

Typically, one uses $D = [0, \kappa_1] \times [0, \kappa_2]$ or
$D=\R_{\geq0}^2$, such that we have $b_1(x,y) = f(x,y) - f(x,0)$ and
$b_2(x,y) = g(x,y) - g(0,y)$, which shows the motivation for this
criterion: 
We compare the population growth of a species in presence of its (potentially
mutualistic) partner against the growth when its partner is absent. 
However, we will see that in cases where $D$ cannot be written as a Cartesian
product, we must use the general definition instead.
In that case, we compare the growth with minimal abundance instead of
absence of the partner (see  \Cref{fig:mutualismcriterion}).
Therefore, our definition reflects the approach typically used in empirical
studies, where all or many partners of the focal species are removed to estimate
the effect of the interaction~\cite{bronstein1994}.
It also resembles the definition for mutualism by~\textcite{demazancourt2005}
using proximate and ultimate response, although we do not discriminate between
genotypes here.

Note that the strong mutualism criterion does not reflect the ecological
characterization of mutualism in cases where the interaction has a positive net
effect for one population, but the benefit is not strictly increasing with its
partner's population size. 
This occurs for instance when the interaction becomes detrimental for large
population sizes of the partner. 
Consider for an example a system with 
$f(x,y) = \rho x(1-x) + \alpha xy(\phi-y)$, in which the benefit for $x$ has a
strict maximum with respect to $y$ and the interaction has a detrimental effect
for $y > \phi$.
With $D=\R_{\geq 0}^2$ and $x>0$ we find
\begin{equation*}
  b_1(x,y) = f(x,y) - f(x,0) = \alpha xy(\phi-y) \geq 0 \iff 0\leq y \leq \phi,
\end{equation*}
as expected, but
\begin{equation*}
  \frac{\partial f}{\partial y}(x,y) = \alpha x(\phi-2y) \geq 0 \iff y \leq
  \frac{1}{2}\phi.
\end{equation*}
Thus, for nonlinear models with interspecific density effects, the strong
mutualism criterion can be too strict and may fail to characterize the
interaction correctly in particular regions of the phase space.

\subsection{Total Population Sizes for System~\eqref{eq:lichenmodel}} 
Models describing the population dynamics of two mutualistic partners are
typically written down as two-dimensional dynamical systems, in which the
state variables represent the population sizes for each species.
For our super model \eqref{eq:lichenmodel}, we consider reductions onto
dimension two as well, but the state variables do not necessarily reflect the
population sizes of host and symbiont.
In order to meaningfully analyse the reductions, we will have to consider the
total population sizes of both partners, which we always denote $X\defl H+C$ for
the mycobiont and $Y\defl S+C$ for the photobiont. 
Since we assumed a 1:1 relation between the two mutualists in the complex, this
represents the total number of individuals of each partner.
Considering System~\eqref{eq:lichenmodel} (and its reductions) with respect to
$X$ and $Y$ is therefore the appropriate way to analyse the effect of the
interaction of one population on the other. 
The corresponding ODE system for total population sizes is always given by
\begin{equation}
  \label{eq:totalpopsystem}
\begin{aligned}
  \frac{dX}{dt} &= \frac{dH}{dt} + \frac{dC}{dt} \\
  \quad \frac{dY}{dt} &= \frac{dS}{dt} + \frac{dC}{dt}
\end{aligned}
\end{equation}
where we have to write the RHS with respect to $X$ and $Y$.
This becomes possible because one of the original components is implicitly
defined by a function in the remaining ones (as defined by the slow manifold).
Due to the fact that $X,Y\geq C$, we often get natural restrictions for the
domain on which System~\eqref{eq:totalpopsystem} can be defined. 
This is also the reason why we have to use the mutualism criterion in
\Cref{def:mutualismcriterion} in the general form.

By considering the total population sizes our reduced systems become comparable
to models used in the literature.
However, due to the nature of the reduction method, it is still helpful and
sometimes necessary to interpret the reductions using the additional information
from the super model.

\section{Reductions for System~\eqref{eq:lichenmodel}
}
\label{sec:reductions}
Applying the method described in  \Cref{sec:findingsfsr} to
System~\eqref{eq:lichenmodel} yields 27 candidates for slow--fast separations of
rates admitting a formal reduction onto a two-dimensional system.
A script for the candidate search and a list with all these TFPVs and their
corresponding reductions can be found in Appendix \ref{apx:reductions} and we
will use the same enumeration in the following.
Here, we consider some notable examples and discuss the general procedure of
computing a reduction.
We will always use the notation in \Cref{def:slow-fast-separation} to denote a
slow--fast separation of rates and write $f^{(0)}=f(\cdot,\pi^\star)$ for the
fast part of System~\eqref{eq:lichenmodel}. 
Then, the affine space in which $\cV(f^{(0)})$ is embedded can be identified
with the phase space. 

\subsection{A Reduction with a Type II Functional Response}
We consider the TFPV candidate 12
\begin{equation*}
  \tilde{\pi} = (\varepsilon \beta_2, \varepsilon \beta_3,
  \delta_1,\varepsilon \delta_2, \varepsilon \delta_3, \varepsilon \mu_1,
  \varepsilon\mu_2, \eta).
\end{equation*}
Then, System~\eqref{eq:lichenmodel} can be written as
\begin{align*}
  f(x, \tilde{\pi}) &= f^{(0)}(x) + \varepsilon f^{(1)}(x) \\ 
                    &=
\begin{pmatrix}
    - \eta S H - \delta_1 H \myphantom \\
    - \eta S H \myphantom \\
    \eta S H  \myphantom
\end{pmatrix} 
  + \varepsilon 
\begin{pmatrix}
    \mu_1 C \left( 1 - \frac{H}{K_1} \right) \\
    \beta_2 S \left( 1 - \frac{S}{K_2}\right) - \delta_2 S + \mu_2 C \left( 1 - \frac{S}{K_2} \right) \\
    \beta_3 C \left( 1 - \frac{C}{K_3}\right) - \delta_3 C 
\end{pmatrix}
\end{align*}
and the affine variety of the fast part $\cV(f^{(0)})=\cV(H)$ has dimension
$s=2$ and no singular point.
This means we can choose $x_0=0$ and find
\begin{equation*}
  Df^{(0)}(x_0) = 
\begin{pmatrix}
    -\delta_1 & 0 & 0 \\
    0 & 0 & 0 \\
    0 & 0 & 0 
\end{pmatrix}
.
\end{equation*}
Thus, Eq.~\eqref{eq:directsum} is satisfied and $\tilde\pi$ is a TFPV for
dimension $2$.
Furthermore, the slow manifold $\cM_0 = \cV(H)$ is attractive if
$\delta_1>0$.

We can choose the product decomposition
\begin{equation*}  
  f^{(0)}(x) = P(x) \psi(x) = 
\begin{pmatrix}
 -\delta_1 -\eta S \\ -\eta S  \\ \eta S 
\end{pmatrix}
 \left(H\right), 
\end{equation*}
and set $\cU_{x_0}\defl \R^3$, so that $\forall x \in
\cU_{x_0}: $ $\text{rank}\, P(x)=\text{rank}\, D\psi(x)=1$ and we have
$\cV(f^{(0)})\cap\cU_{x_0}=\cV(\psi)\cap\cU_{x_0}$.
Thus, we can compute the reduction with~\eqref{eq:reduction}, which ---
upon substituting $H=0$ as defined by the slow manifold to which the reduction
is restricted --- yields the reduced system
\begin{equation}
  \label{eq:reduction12}
\begin{aligned}
    \dot{S} &= \beta_2 S \left(1-\frac{S}{K_2}\right) -  \delta_2 S 
      + \mu_2 C \left(1-\frac{S}{K_2}\right) 
      - \mu_1 C \frac{\eta S}{\delta_1 + \eta S} \\
    \dot{C} &= \beta_3 C \left(1-\frac{C}{K_3}\right) - \delta_3 C 
      + \mu_1 C \frac{\eta S}{\delta_1 + \eta S}
\end{aligned}
\end{equation}

Here, $\mu_1 C$ is the number of autarkic hosts born from the complex,
which are then turned into new lichens whenever $S$ is present. 
This process saturates for increasing $S$, as described by the term 
$\frac{\eta S}{(\delta_1 + \eta S)}$.
Note that this is essentially a type II functional response.
Just as in a predator--prey system, the functional response describes the intake
of organisms depending on the density of their population. 
However, in contrast to the former, the intake is not fatal in this case.
Similar to many predator--prey systems, this reduction is capable of producing
stable oscillations as shown in  \Cref{fig:red12_ts}.

\begin{figure}[t!]
  \centering
  \pgfplotstableread[col sep=comma]{reduction12a_large_eps.csv}\tabA
  \pgfplotstableread[col sep=comma]{reduction12a_small_eps.csv}\tabB
  \pgfplotstableread[col sep=comma]{reduction12b_large_eps.csv}\tabC
  \pgfplotstableread[col sep=comma]{reduction12b_small_eps.csv}\tabD
  \begin{tikzpicture}
    \begin{groupplot}[%
      group style={group size=2 by 2, horizontal sep=10pt, vertical sep=20pt},
      width=0.5\textwidth, height=0.4\textwidth,
      grid=major,
      axis lines=left,
      xmin=0, ymin=0, 
      enlargelimits=true,
      axis background/.style={fill=gray!05}, 
      scaled ticks=false,
      try min ticks=5,
      ]
      \nextgroupplot[
      title={$\varepsilon=\frac{1}{4}$},
      ytick distance={0.3},
      ymax=0.8,
      xticklabel=\empty,
      ]
      \addplot [orange!90!brown, semithick] table [x=t,y=H] {\tabA};
      \addplot [teal!70!green, semithick] table [x=t,y=S] {\tabA};
      \addplot [cyan!40!blue, semithick] table [x=t,y=C] {\tabA};
      \addplot [teal!70!green, thick, densely dotted] table [x=t,y=s] {\tabA};
      \addplot [cyan!40!blue, thick, densely dotted] table [x=t,y=c] {\tabA};
      \coordinate (A) at ({rel axis cs:0,1});
      \nextgroupplot[
      title={$\varepsilon=\frac{1}{100}$},
      xtick distance={500},
      legend entries={
        $H$, $S$, $C$, 
        $\tilde{S}$, $\tilde{C}$
      },
      legend cell align = {left},
      legend style={
        at={(1.04,1)},
        anchor=north west,
        draw=none
      },
      ymax = 0.8,
      xticklabel=\empty,
      yticklabels=\empty,
      ytick distance={0.3}
      ]
      \addplot [orange!90!brown, semithick] table [x=t,y=H] {\tabB};
      \addplot [teal!70!green, semithick] table [x=t,y=S] {\tabB};
      \addplot [cyan!40!blue, semithick] table [x=t,y=C] {\tabB};
      \addplot [teal!70!green, thick, densely dotted] table [x=t,y=s] {\tabB};
      \addplot [cyan!40!blue, thick, densely dotted] table [x=t,y=c] {\tabB};
      \coordinate (B) at ({rel axis cs:0,1});
      \nextgroupplot[
      ymax = 0.52,
      xlabel=$t$,
      ytick distance={0.2}
      ]
      \addplot [orange!90!brown, semithick, smooth] table [x=t,y=H] {\tabC};
      \addplot [teal!70!green, semithick, smooth] table [x=t,y=S] {\tabC};
      \addplot [cyan!40!blue, semithick, smooth] table [x=t,y=C] {\tabC};
      \addplot [teal!70!green, thick, densely dotted, smooth] table [x=t,y=s] {\tabC};
      \addplot [cyan!40!blue, thick, densely dotted, smooth] table [x=t,y=c] {\tabC};
      \coordinate (C) at ({rel axis cs:0,1});
      \nextgroupplot[
      xlabel=$t$,
      ymax = 0.52,
      ytick distance={0.2},
      yticklabels=\empty,
      ]
      \addplot [orange!90!brown, semithick, smooth] table [x=t,y=H] {\tabD};
      \addplot [teal!70!green, semithick, smooth] table [x=t,y=S] {\tabD};
      \addplot [cyan!40!blue, semithick, smooth] table [x=t,y=C] {\tabD};
      \addplot [teal!70!green, thick, densely dotted, smooth] table [x=t,y=s] {\tabD};
      \addplot [cyan!40!blue, thick, densely dotted, smooth] table [x=t,y=c] {\tabD};
      \coordinate (D) at ({rel axis cs:0,1});
    \end{groupplot}
    \node[anchor=south west] at (A) {\small \textbf{a}};
    \node[anchor=south west] at (B) {\small \textbf{b}};
    \node[anchor=south west] at (C) {\small \textbf{c}};
    \node[anchor=south west] at (D) {\small \textbf{d}};
  \end{tikzpicture}
  \caption{Simulations for System~\eqref{eq:reduction12} with $\pi=\tilde{\pi}$ for
    varying $\varepsilon$ and base parameters:
    \textbf{a}, \textbf{b}: 
    $\beta_2 = 4$, $\beta_3 = 2$, $\delta_1 = 5$, $\delta_2 = 1$,
    $\delta_3 = 1.5$, $\mu_1 = 3$, $\mu_2 = 0.5$, $\eta = 10$, 
    \textbf{c}, \textbf{d}:
    $\beta_2 = 6$, $\beta_3 = 3$, $\delta_1 = 1$, $\delta_2 = 1$,
    $\delta_3 = 6$, $\mu_1 = 6$, $\mu_2 = 1$, $\eta = 12$.
    \textbf{a}, \textbf{c}: 
    $\varepsilon=\frac{1}{4}$, 
    \textbf{b}, \textbf{d}:
    $\varepsilon=\frac{1}{100}$.
    The resource defined capacities are $K_1 = K_2 = K_3 = 1$ and the initial
    values are $H(0) = 0.01, S(0) = C(0) = 0.08$ in each case.
    Note that for small values of $\varepsilon$ the reduced system becomes a
    very good approximation of the original system.
  }
  \label{fig:red12_ts}
\end{figure}

To review the net effect of the interaction on the autarkic symbiont population,
we can consider all effects of the interaction:
Whilst the population gets depleted with a type II response, the interaction
also results in births into the autarkic symbiont population (given that there
are enough resources for a successful establishment). 
The interaction becomes detrimental for the autarkic population $S$ whenever it
exceeds the critical population size
\begin{equation*}
  S_{\text{c}} \defl
  \frac{\eta K_2 \left( \mu_2 - \mu_1 \right) - \delta_1 \mu_2}{2 \eta \mu_2} +
  \sqrt{\frac{\left( \eta K_2 \left( \mu_2 - \mu_1 \right) - \delta_1 \mu_2
    \right)^2}{4 \eta^{2} \mu_2^{2}}
  + \frac{K_2 \delta_1}{\eta}},
\end{equation*}
which is always positive.
However, even if the effect of the interaction on the symbiont is negative, the
overall effect for the corresponding population need not be, since we have to
account for autarkic and mutualistic individuals.
To see this more clearly, we can rewrite System~\eqref{eq:reduction12} for the total
populations of host and symbiont $X$ and $Y$, respectively, as in
\eqref{eq:totalpopsystem}. 
In this case, we have $H=0$ on the slow manifold, which implies $\dot{H}=0$ as
well as $C=X$ and $S=Y-X$. 
With this, System~\eqref{eq:reduction12} for total population sizes is
\begin{equation}
  \label{eq:reduction12_total}
\begin{aligned}
    \dot{X} ={}& 
    \beta_3 X \left( 1 - \frac{X}{K_3} \right) - \delta_3 X
    + \frac{\eta \mu_1 X (Y-X)}{\delta_1 + \eta (Y-X) } \\ 
    \dot{Y} ={}& 
    \beta_3 X \left( 1 - \frac{X}{K_3} \right) - \delta_3 X  \\
               & \quad
    + \left(\beta_2 \left( Y-X \right) + \mu_2 X \right)\left( 1 - \frac{Y - X}{K_2} \right) - \delta_2 \left( Y - X \right) 
\end{aligned}
\end{equation}
which only makes sense on the domain 
$D=\set{(X,Y)\in\R_{\geq0}^2 \mid X \leq Y}$ due to the fact that
$Y-X=S\geq0$. 
Applying the mutualism criterion, i.e.\ \Cref{def:mutualismcriterion}, to
System~\eqref{eq:reduction12_total} on $D$ yields the benefit functions
\begin{align*}
  b_1(X,Y) ={}& \frac{\eta \mu_1 X (Y - X)}{\delta_1 + \eta (Y - X)}
    \\
  b_2(X,Y) ={}& 
    \frac{\beta_2 Y^2 - \left(Y - X \right) \left(\beta_2 (Y - X) +
    \mu_2 X \right) }{K_2} \\
            & \quad + \left( \beta_3 \left( 1 - \frac{X}{K_3} \right) - \delta_3 - (\beta_2 - \delta_2) + \mu_2 \right)X
\end{align*}
\begin{figure}[t]
  \centering
  \includegraphics[width=411.349pt]{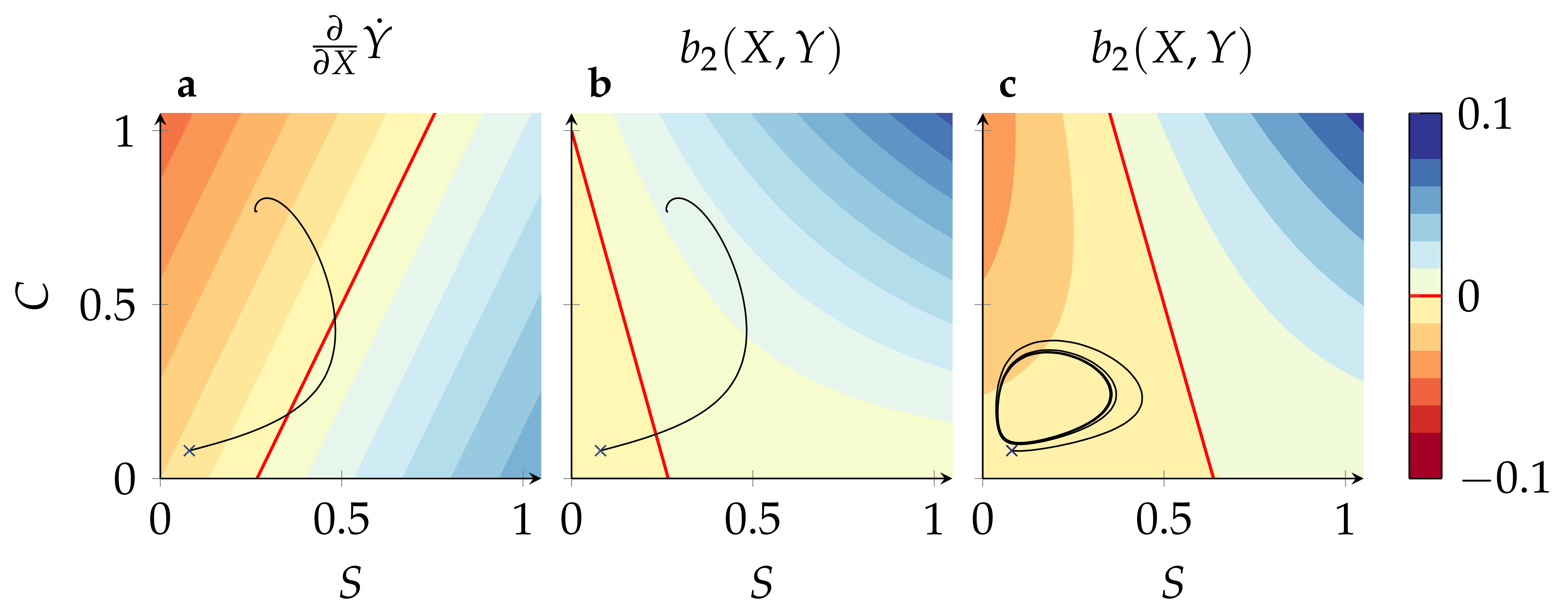}
  \caption{\textbf{a}, \textbf{b}: Strong mutualism criterion
    (\Cref{def:strongmutualismcriterion}) and benefit function $b_2(X,Y)$
    (\Cref{def:mutualismcriterion}) measuring the effect of the interaction on
    $Y$ in System~\eqref{eq:reduction12_total} with $\pi=\tilde{\pi}$ and
    $\varepsilon$ as in \Cref{fig:red12_ts}~b.
    \textbf{c}: Benefit function $b_2(X,Y)$ for parameters as in
    \Cref{fig:red12_ts}~d. 
    The black line shows the simulated trajectories as in the corresponding
    plots in \Cref{fig:red12_ts} (the initial condition is marked).
    Note, that the two mutualism criteria differ for the first scenario: 
    According to the strong mutualism criterion the system is not mutualistic
    in the steady state, but the benefit functions suggest that it is.
  }
  \label{fig:red12_contour}
\end{figure}
Therefore, the mycobiont always has an advantage, whilst the situation is
more complex for the photobiont.
Note that we can substitute back to obtain $b_1$ and $b_2$ with respect to $S$
and $C$.
The net effect of the interaction for the photobiont measured with the benefit
function is shown in  \Cref{fig:red12_contour} for the two parameter sets used in
 \Cref{fig:red12_ts}, together with the partial derivative
$\frac{\partial}{\partial X} \dot{Y}$, i.e.\ the strong mutualism criterion for
the symbiont.
We can observe, that the strong mutualism criterion does not indicate that the
system is mutualistic in the first scenario along most of the trajectory.
However, the benefit function for the photobiont is positive after a short
initial phase. 
Most importantly, both criteria disagree in the stable fixed point.
For the second scenario, the benefit function $b_2$ becomes negative in the
relevant region. 
Thus, System~\eqref{eq:reduction12} is able to depict shifts from a mutualistic
interaction to a parasitic relation with the typical oscillatory behaviour of
predator--prey systems.

\begin{figure}[t]
  \pgfplotstableread[col sep=comma]{reduction12_quiver.csv}\quiver
  \pgfplotstableread[col sep=comma]{reduction12_S_isoclines.csv}\isoS
  \pgfplotstableread[col sep=comma]{reduction12_C_isoclines.csv}\isoC
  \pgfplotstableread[col sep=comma]{reduction12_separatrix.csv}\separatrix
  \begin{center}
    \begin{tikzpicture}
      \begin{axis}[
        width=0.75\textwidth, height=0.5\textwidth,
        xlabel=$S$, ylabel=$C$,
        legend cell align={left},
        legend style={
          at={(1.04,1)},
          anchor=north west,
          draw=none
        },
        axis lines=left,
        xmin=-0.1, xmax=8.0,
        ymin=-0.1, ymax=4.0,
        restrict y to domain=-1:7,
        enlargelimits=false,
        axis background/.style={fill=gray!05}
        ]
        \addplot [
          blue,
          quiver={u=\thisrow{u},v=\thisrow{v}},
          quiver/colored = {mapped color},
          point meta=\thisrow{c},
          -stealth,
          colormap name={viridis},
          ] table \quiver;
        \addplot [thick, orange!90!brown] table [x=x1, y=y1] \isoS;
        \addplot [thick, teal!70!green] table [x=x1, y=y1] \isoC;
        \addplot [thick, teal!70!green] table [x=x2, y=y2] \isoC;
        \addplot[thick, gray] table [x=S, y=C] \separatrix;
        \addplot [
          scatter,
          only marks,
          point meta=explicit symbolic,
          scatter/classes={
            stable={mark=*,draw=black,fill=gray},
            unstable={mark=*,draw=black,fill=white} 
          },
          ] table [meta=stability] { 
            x y stability 
            0.0 0.0 unstable 
            0.29235128730684706 1.6407748856286453 stable
            1.913048598617692 2.9389584362720704 unstable
            4.024241059525439 3.1360592345392915 stable 
            7.5 0.0 unstable
          };
        \legend{, $\dot{S}=0$, $\dot{C}=0$, , separatrix, stable, unstable}
      \end{axis}
    \end{tikzpicture} 
  \end{center}
  \caption{Vector field of System~\eqref{eq:reduction12} with nullclines given
    in~\eqref{eq:red12nullclines},
    the biologically relevant stable and unstable fixed points and the
    separatrix (computed numerically), which defines the basins of
    attraction for the stable positive fixed points, for parameters
    $\beta_2 = 8, \beta_3 = 6, \delta_1 = 2, \delta_2 = 2, \delta_3 = 7, \mu_1 =
  5, \mu_2= 2, \eta = 10, K_1 = 10, K_2 = 10, K_3 = 5$.\label{fig:red12_bistability}}
\end{figure}

The nullclines for System~\eqref{eq:reduction12} are given by
\begin{equation}
  \label{eq:red12nullclines}
\begin{aligned}
    \dot{S} = 0 &\iff C = \frac{\beta_2\eta S 
      \left(S - \frac{\beta_2-\delta_2}{\beta_2} K_2\right)
      \left(S + \frac{\delta_1}{\eta}\right)}
      {\mu_2(K_2 - S)(\delta_1 + \eta S) - \mu_1 \eta K_2 S} \\ 
    \dot{C} = 0 &\iff C =  \left(\beta_3 - \delta_3 + \mu_1 \frac{\eta S}{\delta_1 + \eta S}\right) \frac{K_3}{\beta_3}
      \quad\text{or}\quad C = 0
\end{aligned}
\end{equation}
Thus, there exist three trivial fixed points for $C=0$, of which $(0,0)$ and
$(\frac{\beta_2-\delta_2}{\beta_2} K_2, 0)$ are biologically relevant, and at
most four nontrivial fixed points with $C\neq0$, since equating both nullclines
yields a fourth order polynomial in $S$ that vanishes if and only if the
nullclines intersect.

Just as the original system, this reduction may show bistability (with multiple
stable interior fixed points), which can be seen in  \Cref{fig:red12_bistability}.
In the scenario depicted, the total population size of photobionts $S+C$ is
smaller in both stable positive fixed points compared to their carrying capacity
$\frac{\beta_2-\delta_2}{\beta_2}K_2$ attained in the autarkic trivial fixed
point, which indicates that they do not benefit from the formation of lichens.
Evaluating the mutualism criterion for the parameters as in
 \Cref{fig:red12_bistability} shows indeed that in both points the net benefit
for the photobiont is negative, but the net benefits are larger for both
partners in the stable fixed point on the upper right. 

The location of the separatrix implies that the initial population size of the
complex is most important for the long time behaviour of the system in the sense
that for small $C(0)$ the system always converges to the fixed point with higher
net benefit.
Biologically speaking, this bistability arises from the negative effect the
photobiont experiences due to the interaction, which ultimately is due to the
negative net growth of the complex, i.e.\ $\beta_3 < \delta_3$.
If the initial lichen population is large, more photobionts are incorporated
into new lichens and the autarkic photobiont population is not able to offset 
the loss in the complex.
If on the other hand the autarkic population of photobionts is sufficiently
large or the lichen population is small, the higher number of photobionts can
support significantly more hosts and lichens resulting in a higher net benefit
for both partners.

\subsection{A Reduction with Multiple Slow Manifolds}
\label{sec:reduction1}
We consider the TFPV candidate 1
\begin{equation*}
  \tilde{\pi} = (\varepsilon \beta_2, \varepsilon \beta_3, \varepsilon
  \delta_1,\varepsilon \delta_2, \varepsilon \delta_3, \varepsilon \mu_1,
  \varepsilon\mu_2, \eta).
\end{equation*}
In this case, all rates except the formation rate of the complex $\eta$ are
small.
Then, the affine variety of the fast part consists of two irreducible components
\begin{equation*}
  \cV(f^{(0)}) = 
  \cV(SH) = 
      \cV(S) \cup \cV(H)
  \defr Y_1 \cup Y_2.
\end{equation*}
Each has dimension $s=2$ and admits a reduction. 

\subsubsection{Reduction (a): \texorpdfstring{$\cM_0 = Y_1$}{M\_0 $=$ Y\_1}}
\label{sec:reduction1i}
The point $x_0=(H',0,0)$, for arbitrary $H'>0$, is a nonsingular point of
$Y_1$. 
The Jacobian of $f^{(0)}$ at $x_0$ is given by
\begin{equation*} 
  Df^{(0)}(x_0) =
\begin{pmatrix}
    0 &  - \eta H' & 0 \\
    0 &  - \eta H' & 0 \\
    0 & \eta H' & 0 
\end{pmatrix}
,
\end{equation*}
has rank $r=1$ as required and satisfies~\eqref{eq:directsum}.
From \Cref{thm:reduction} follows that $\tilde{\pi}$ is indeed a TFPV for
dimension 2 with slow manifold $\cM_0 = Y_1$.

To compute the reduced system, we can choose the product decomposition
\begin{equation*} 
  f^{(0)}(x) = P(x)\psi(x) = 
\begin{pmatrix}
    -H \\ -H \\ H
\end{pmatrix}
 (\eta S).
\end{equation*}
Then, $\cV(\psi) = Y_1$ is satisfied and with the open Zariski neighbourhood
$\cU_{x_0} \defl \R^3\setminus Y_2$ follows $\forall x\in \cU_{x_0}: $
$\text{rank}\; P(x) = 1$ and $\cV(f^{(0)})\cap \cU_{x_0} = \cV(\psi) \cap
\cU_{x_0}$, which means $P$ and $\psi$ satisfy the requirements in
\ref{thm:reduction-composition} of \Cref{thm:reduction}. 
The reduced system is then given by
\begin{equation}
  \label{eq:reduction1a}
\begin{aligned}
    \dot{H} &= \mu_1 C \left( 1 - \frac{H}{K_1} \right) - \delta_1 H -  \mu_2 C  \\ 
    \dot{C} &= \beta_3 C \left( 1 - \frac{C}{K_3}\right) - \delta_3 C   +  \mu_2 C
\end{aligned}
\end{equation}
The only nonzero eigenvalue of $Df^{(0)}$ is $-\eta H' < 0$ and therefore the
slow manifold $Y_1$ is attractive on some time interval due to
\ref{thm:reduction3} in \Cref{thm:reduction}.
Note that solutions may still leave $\cM_0$ after some time. 
Moreover, since the nonzero eigenvalue of $Df^{(0)}$ depends on $H'$, the slow
manifold loses its stability if $H'$ tends to zero.

\begin{remark}
This situation shows the typical procedure if $\cV(f^{(0)})$ has
  multiple irreducible components. 
  In particular, the general idea is to set
  \begin{equation*} 
    \cU_{x_0}
    = \R^n \setminus \bigcup_{i=1}^k Y_i , 
\end{equation*}
  where $\cV(f^{(0)}) = \bigcup_{i=0}^k Y_i$ such that $Y_i$ are the
  irreducible components of $\cV(f^{(0)})$ and $x_0\in Y_0$. 
  Then, we can use the generators of $Y_0$ as entries of $\psi$, which directly
  implies $\cV(\psi)=Y_0$ and thus
  $\cV(\psi)\cap\cU_{x_0} = 
  \cV(f^{(0)})\cap\cU_{x_0}$ is always satisfied.
\end{remark}

We can rewrite System~\eqref{eq:reduction1a} as in \eqref{eq:totalpopsystem} by substituting
$C = Y$ and $H = X - Y$ according to the slow manifold. 
The net benefits as in \Cref{def:mutualismcriterion} on 
$D=\set{(X,Y)\in\R_{\geq0}^2 \mid Y \leq X}$ are
\begin{equation*}
  b_1(X,Y) = \frac{\mu_1(Y-X)Y}{K_1} + \left(\beta_3\left(1-\frac{Y}{K_3}\right) -
  \delta_3 + \delta_1 + \mu_1\right) Y
\end{equation*}
and $b_2 \equiv 0$.
The lower bound for $X$ comes from the fact that the system only makes sense
for $X-Y=H\geq 0$.
Thus, the reduction describes commensalism on
\[
  D^+ = \left\{
    (x,y)\in\R_{\geq0}^2 \midvert 
    y \leq x \leq \left(1 - \frac{\beta_3 K_1}{\mu_1 K_3}\right) y  + K_1\left(1 + \frac{\beta_3 - \delta_3 - \delta_1}{\mu_1}\right)
  \right\}
\]
and amensalism on $D\setminus D^+$.

\subsubsection{Reduction (b): \texorpdfstring{$\cM_0 = Y_2$}{M\_0 $=$ Y\_2}}
\label{sec:reduction1ii}
Now we can choose $x_0=(0,S',0)$ for some $S'>0$, which is a nonsingular point
of the component $Y_2$.
All steps required to compute the corresponding reduction are analogous to the
previous case.
The reduced system is then given by
\begin{equation}
\label{eq:reduction1b}
\begin{aligned}
 \dot{S} &= \beta_2 S \left(1-\frac{S}{K_2}\right) - \delta_2 S + \mu_2 C \left(1-\frac{S}{K_2}\right) - \mu_1 C  \\
 \dot{C} &= \beta_3 C \left(1-\frac{C}{K_3}\right) -\delta_3 C + \mu_1 C
\end{aligned}
\end{equation}
Just as before, the slow manifold $Y_2$ is locally attractive because the only
nonzero eigenvalue of $Df^{(0)}(x_0)$ is $-\eta S' <  0$.

Again, using total population sizes by substituting $C=X$ and $S=Y-X$ in
\eqref{eq:reduction1b} and rewriting the system as in \eqref{eq:totalpopsystem}
allows us to apply the mutualism criterion for
$D=\set{(X,Y)\in\R^2\mid X \leq Y}$. 
This shows that the interaction is neutral for the host and the symbiont has a
positive net benefit if $X>0$ and
\[
  2\beta_2 \gtrless \mu_2
  \quad \text{and} \quad
  Y \gtrless \frac{\left(\beta_2 - \mu_2 + \beta_3 \frac{K_2}{K_3} \right) X +
  K_2 \left(\beta_2 - \delta_2  - \mu_2 - (\beta_3 - \delta_3)\right)}{2 \beta_2
- \mu_2}.
\]

\subsubsection{Behaviour of the Reductions near Singular Points}
First we note that in both reduced systems the complex $C$ does not depend on
the autarkic population and simply grows logistically (with increased
per-capita birth rate compared to the original system).
This fact renders both models not particularly interesting,
but they allow us to demonstrate a property of the convergence in Tikhonov's
theorem: Namely that it can only be guaranteed on some, possibly finite,
time interval.
One reason for this is that solutions may enter unstable regions of the slow
manifold, as we can observe here.

\begin{figure}[t!]
  \pgfplotstableread[col sep=comma]{reduction1.csv}\tsRedA
  \centering
  \begin{tikzpicture}
    \begin{axis}[
      xlabel=$t$,
      width=0.75\textwidth, height=0.5\textwidth,
      legend entries={$H$,$S$,$C$,$\tilde{H}_a$,$\tilde{C}_a$,$\tilde{S}_b$, $\tilde{C}_b$},
      legend cell align={left},
      legend style={
        at={(1.04,1)},
        anchor=north west,
        draw=none
      },
      grid=major,
      xtick={0,50,100,150,200},
      axis lines=left,
      xmin=0, ymin=0, ymax=1.4,
      enlargelimits=true,
      axis background/.style={fill=gray!05}
      ]
      \addplot [orange!90!brown, semithick] table [x=t,y=H] {\tsRedA};
      \addplot [teal!70!green, semithick] table [x=t,y=S] {\tsRedA};
      \addplot [cyan!40!blue, semithick] table [x=t,y=C] {\tsRedA};
      \addplot [orange!90!brown, thick, densely dotted] table [x=t,y=H1a] {\tsRedA};
      \addplot [cyan!40!blue, thick, densely dotted] table [x=t,y=C1a] {\tsRedA};
      \addplot [teal!70!green, thick, densely dashed] table [x=t,y=S1b] {\tsRedA};
      \addplot [cyan!40!blue, thick, densely dashed] table [x=t,y=C1b] {\tsRedA};
    \end{axis}
  \end{tikzpicture}
  \caption{Simulated time series for System~\eqref{eq:lichenmodel} and the
    reduction \eqref{eq:reduction1a} with components denoted $\tilde{H}_a$ and $\tilde{C}_a$
    as well as 
    reduction \eqref{eq:reduction1b} with components denoted $\tilde{S}_b$ and $\tilde{C}_b$.
    The simulations are obtained with $\varepsilon = 0.01$ and base parameters 
    $\beta_2 = 4$, $\beta_3 = 3$, $\delta_1 = 1$, $\delta_2 = 3$, $\delta_3 =
    1$, $\mu_1 = 2$, $\mu_2 = 4$, $\eta = 5$ and $K_1 = K_2 = K_3 = 1$, the
    initial values are $H(0)=C(0)=0.5$ and $S(0)=0.005$.
  }
  \label{fig:reduction1}
\end{figure}

For System~\eqref{eq:reduction1a}, $C(0)>0$ and $\beta_3-\delta_3+\mu_2>0$
imply
\begin{equation*}
  \lim_{t\to\infty} C(t) = \frac{\beta_3-\delta_3+\mu_2}{\beta_3}K_3.
\end{equation*}
In this case, $H$ becomes negative if additionally $\mu_1<\mu_2$.
This behaviour is similar in System~\eqref{eq:reduction1b}: $S$ becomes negative if
$\beta_2<\delta_2$ and $\mu_1>\mu_2$.
Because \Cref{thm:solutionbounded} guarantees that all components of
System~\eqref{eq:lichenmodel} remain nonnegative,
deciding which reduction is a good approximation for the long time behaviour of
the super model depends not only on the initial condition but also on the
parameters $\mu_1$ and $\mu_2$. 
\Cref{thm:reduction} does not guarantee the existence of a two-dimensional
reduction along the $C$-axis, since it consists of singular points in which the
reductions is invalid (\Cref{thm:reduction} requires that $x_0$ is nonsingular).

Solutions starting near one irreducible component may approach another
after finite time and follow a different reduction.
 \Cref{fig:reduction1} illustrates this behaviour: 
The initial population sizes are positive and near $Y_1$, but $\mu_1 < \mu_2$
and thus the corresponding reduction, i.e.\ System~\eqref{eq:reduction1a}, is
only a good approximation for the initial phase of the dynamics. 
The solution leaves the slow manifold $Y_1$, because the latter loses
stability when $H$ passes zero and becomes negative.
The long time behaviour is much better described by System~\eqref{eq:reduction1b}, to
which the full system converges for large times. 

Thus, whenever there exist more than one slow manifold, solutions may approach
their intersection, which are singular points of the affine variety
$\cV(f^{(0)})$. 
Since the reduction method is not applicable in these points,
it is not clear how the dynamics evolve within them. 
Here, the autarkic populations can become negative in both reductions and we can
see that the corresponding slow manifolds can become unstable due to the fact
that they are only locally attractive (both slow manifolds are attractive if the
respective autarkic population remains positive).

\subsection{A Reduction Towards a Predator--Prey System}
\label{sec:reduction20}
The TFPV candidate 20
\begin{equation*}
  \tilde{\pi} = \left( \varepsilon\beta_2, \beta_3, \varepsilon\delta_1,
  \varepsilon\delta_2, \delta_3, \mu_1, \varepsilon\mu_2, \varepsilon\eta \right)
\end{equation*}
admits the affine variety
\begin{equation*}
  \cV(f^{(0)}) = 
     \cV(C) \cup \cV\left(\beta_3 \left(1-\frac{C}{K_3}\right) - \delta_3, H - K_1\right)
          \defr Y_1 \cup Y_2,
\end{equation*}
with only the first irreducible component $Y_1$ having dimension $s=2$. 
For the nonsingular point $x_0=0\in Y_1$ we find
\begin{equation*}
  D f^{(0)} (x_0) = 
\begin{pmatrix}
    0 & 0 & \mu_1 \\
    0 & 0 & 0 \\
    0 & 0 & \beta_3 - \delta_3 
\end{pmatrix}
,
\end{equation*}
which means that $\tilde{\pi}$ is a TFPV for dimension 2 and the reduced system
is attractive if $\beta_3<\delta_3$. 
A product decomposition for the fast part satisfying the conditions in
\Cref{thm:reduction} is
\begin{equation*}
  f^{(0)}(x) = P(x) \psi(x) = 
\begin{pmatrix}
    \mu_1\left(1 - \frac{H}{K_1}\right) \\
    0 \\
    \beta_3 \left(1 - \frac{C}{K_3}\right) - \delta_3
\end{pmatrix}
  \left(C\right) .
\end{equation*}
This yields the reduced system
\begin{equation}
\begin{aligned}
  \dot{H} &=  - \delta_1 H - \eta \left(1 +  \frac{\mu_1}{\beta_3 - \delta_3} \left( 1 - \frac{H}{K_1} \right) \right) S H \\
                    \dot{S} &= \beta_2 S \left( 1 - \frac{S}{K_2} \right) - \delta_2 S - \eta S H
\end{aligned}
\end{equation}
Note that this is only a good approximation of System~\eqref{eq:lichenmodel} if
$\beta_3<\delta_3$. 
Since $C=0$, we can measure the benefit of both partners directly by considering
the effects on the autarkic populations.
Then, given $0\leq H \leq K_1$, the host has a net benefit from the interaction
if and only if $\beta_3-\delta_3+\mu_1>0$, which we will assume in the
following.
We define
\begin{equation}
  \label{eq:substitutionpredprey}
  \gamma_i \defl - \frac{\beta_3 - \delta_3 + \mu_i}{\beta_3 - \delta_3}
  \quad\text{and}\quad
  L_i \defl \frac{\beta_3 - \delta_3 + \mu_i}{\mu_i} K_i,
\end{equation}
for $i=1,2$ (we will use these parameters again later).
Given $\beta_3<\delta_3$ and $\beta_3-\delta_3+\mu_i>0$, we find $\gamma_i>0$
and $0<L_i<K_1$. 
These definitions seem biologically reasonable, since it means the rescaled
per-capita birth rate $\gamma_i$ is large if the mutualistic birth rate $\mu_i$
itself is large (births occur quickly from the interaction) or
$\abs{\beta_3-\delta_3}$ is small (more time for reproduction since the interaction
lasts longer). 
Moreover, $\gamma_i=0$ exactly if $\mu_i=-(\beta_3-\delta_3)$, i.e.\ if births
from the complex occur with the same rate as its population vanishes.
The rescaled capacity $L_i$ reflects the dynamic carrying capacity, i.e.\ the
stable population size depending on resource availability, births and deaths.
It is therefore also sensible that it increases with $\mu_i$.

With the definitions above, we can rewrite the reduced system as
\begin{equation}
  \label{eq:reduction20}
\begin{aligned}
    \dot{H} &= - \delta_1 H + \gamma_1 \left( 1 - \frac{H}{L_1} \right) \eta S H \\
    \dot{S} &= \beta_2 S \left( 1 - \frac{S}{K_2} \right) - \delta_2 S - \eta S H
\end{aligned}
\end{equation}
Now it became obvious that this reduction resembles a predator--prey system: 
$H$ gains a benefit from the interaction, while $S$ only experiences negative
effects (we can see this with the benefit functions for System~\eqref{eq:reduction20}).
Here, $\eta S$ is the functional response of type I. 
The numerical response of the predator $H$ is density dependent with logistic
conversion rate.
As in a predator--prey system, we can think of the term $\eta S H$ as births of
the predator enabled by the interaction, but due to the limited availability of
resources not all births are successful --- hence the logistic conversion rate. 
In terms of the lichen symbiosis this means the mycobiont reproduces with a rate
proportional to the probability of an individual encountering a photobiont, but
the successful establishment depends on its population density. 

This reduction suggests that mutualism can break down completely when the
benefit becomes asymmetrical, which is known to occur in real systems
\cite{bronstein2001}.
For this TFPV we have $\mu_1 \gg \varepsilon\mu_2$, which results in a reduced
system in which only $H$ experiences a benefit in a first order approximation.
Thus, albeit this is not obvious from the model formulation of
System~\eqref{eq:lichenmodel}, we have seen that our super model is capable of
describing the shift of the interaction towards parasitism.

Note that although we did not include costs of the mutualism explicitly in the
model derivation, they became apparent in this reduction. 
While it is clear that the benefit for $S$ vanishes due to $\varepsilon\mu_2$
being small, it is perhaps surprising that we can now see the costs associated
with the mutualistic interaction as the term $-\eta S H$.
In this scenario, the complex $C$ essentially functions as a catalyst or
intermediate state corresponding to the predation process: $S$ and $H$ form
$C$, which then immediately breaks down to $H$ due to the asymmetrical benefits.
Thus, the symbiont no longer gains any reproductive advantage while still
experiencing the costs associated to the interaction.

\subsection{A Reduction with Logistic Benefits}
The TFPV candidate 21
\begin{equation*}
  \tilde{\pi} = (\varepsilon\beta_2, \beta_3, \varepsilon\delta_1,
  \varepsilon\delta_2, \delta_3, \mu_1, \mu_2, \varepsilon\eta)
\end{equation*}
admits the affine variety
\begin{equation*}
  \cV(f^{(0)}) 
  = \cV(C) \cup \cV\left(\beta_3 \left(1-\frac{C}{K_3}\right) - \delta_3, H - K_1, S - K_2\right)
  \defr Y_1 \cup Y_2.
\end{equation*}
However, only the irreducible component $Y_1$ has dimension $s=2$ as desired.
We can choose the nonsingular point $x_0=0\in Y_1$. 
Then, we find
\begin{equation*}
  D f^{(0)}(x_0) = 
\begin{pmatrix}
    0 & 0 & \mu_1 \\
    0 & 0 & \mu_2 \\
    0 & 0 & \beta_3-\delta_3
\end{pmatrix}
\end{equation*}
and it follows that $\tilde{\pi}$ is a TFPV for dimension 2.
The slow manifold $Y_1$ is attractive if $\beta_3<\delta_3$ and
\[
  f^{(0)}(x) = P(x)\psi(x) = 
  \begin{pmatrix}
    \mu_1 \left(1-\frac{H}{K_1}\right) \\ 
    \mu_2 \left(1-\frac{S}{K_2}\right) \\
    \beta_3 \left(1-\frac{C}{K_3}\right) - \delta_3 
  \end{pmatrix} (C)
\]
is a possible product decomposition of $f^{(0)}$.
With $\cU_{x_0} \defl \R^3\setminus Y_2$ we can see that $P$ and $\psi$
satisfy all requirements in \Cref{thm:reduction}
\ref{thm:reduction-composition}. 
Using the substitutions \eqref{eq:substitutionpredprey} and with
$\tilde{\gamma}_i \defl \eta \gamma_i$ for $i=1,2$, 
the corresponding reduction is
\begin{equation}
  \label{eq:reduction21_rewrite}
\begin{aligned}
    \dot{H} &= -\delta_1 H + \tilde{\gamma}_1 \left(1-\frac{H}{L_1}\right) S H  \\
    \dot{S} &= \beta_2 S \left(1-\frac{S}{K_2}\right) - \delta_2 S + 
    \tilde{\gamma}_2 \left(1-\frac{S}{L_2}\right) S H 
\end{aligned}
\end{equation}

The benefit functions for this system are
\begin{equation*}
  b_1(H,S) = \tilde{\gamma}_1 \left(1-\frac{H}{L_1}\right) S H 
  \quad\text{and}\quad 
  b_2(H,S) =\tilde{\gamma}_2 \left(1-\frac{S}{L_2}\right) S H.
\end{equation*}
Since we assumed $\beta_3<\delta_3$ (for the slow manifold to be attractive),
the net benefits are positive if $\beta_3-\delta_3+\mu_i > 0$ and $H<L_1$ and 
$S<L_2$.
Thus, in contrast to System~\eqref{eq:reduction20}, this reduction actually
describes mutualism on the domain $[0, L_1]\times [0,L_2]$ according to the
mutualism criterion, which is the result of $\mu_2$ not being a small parameter
as was the case for the previous TFPV.
Note that for $H(0)\in \opint{0,L_1}$ we find $H(t)<L_1$, since
$\dot{H}\vert_{H=L_1} = -\delta_1 L_1 < 0$. 
Furthermore,
\begin{equation*}
  \dot{S}\vert_{S=L_2} = -\left(\beta_2\frac{\beta_3-\delta_3}{\mu_2} + \delta_2\right)L_2,
\end{equation*}
which implies that $S > L_2$ is possible if $\beta_2-\delta_2$ is large, $\mu_2$
is small or $\delta_3-\beta_3$ is large. 
In this case, the benefit of additional births is not sufficient to offset the
costs associated to producing $C$, since the autarkic growth is more effective
for the symbiont compared to the mutualistic growth. 

\begin{figure}[t]
  \centering
  \pgfplotstableread[col sep=comma]{reduction21_quiver.csv}\quiver
  \pgfplotstableread[col sep=comma]{reduction21_H_isoclines.csv}\isoH
  \pgfplotstableread[col sep=comma]{reduction21_S_isoclines.csv}\isoS
  \pgfplotstableread[col sep=comma]{reduction21_separatrix.csv}\separatrix
  \begin{tikzpicture}
    \begin{axis}[
      width=0.75\textwidth, height=0.5\textwidth,
      xlabel=$H$, ylabel=$S$,
      legend cell align={left},
      legend style={
        at={(1.04,1)},
        anchor=north west,
        draw=none
      },
      axis lines=left,
      xmin=-0.06, xmax=0.76,
      ymin=-0.06, ymax=1.11,
      restrict y to domain=-1:2,
      enlargelimits=false,
      axis background/.style={fill=gray!05}
      ]
      \addplot [
        blue,
        quiver={u=\thisrow{u},v=\thisrow{v}},
        quiver/colored = {mapped color},
        point meta=\thisrow{c},
        -stealth,
        colormap name={viridis},
        ] table \quiver;
      \addplot [thick, orange!90!brown] table [x=x1, y=y1] \isoH;
      \addplot [thick, orange!90!brown] table [x=x2, y=y2] \isoH;
      \addplot [thick, teal!70!green] table [x=x1, y=y1] \isoS;
      \addplot [thick, teal!70!green] table [x=x2, y=y2] \isoS;
      \addplot[thick, gray] table [x=H, y=S] \separatrix;
      \addplot [
        scatter,
        only marks,
        point meta=explicit symbolic,
        scatter/classes={
          stable={mark=*,draw=black,fill=gray},
          unstable={mark=*,draw=black,fill=white} 
        },
        ] table [meta=stability] { 
          x y stability 
          0.0 0.0 stable 
          0.5947710159851851 0.8637091960510017 stable 
          0.12189565068148153 0.39858588591621136 unstable 
        };
      \legend{,$\dot{H}=0$, , $\dot{S}=0$, , separatrix, stable, unstable}
    \end{axis}
  \end{tikzpicture} 
  \caption{Vector field defined by
    System~\protect\eqref{eq:reduction21_rewrite} with zero-isoclines and
    stable and unstable equilibrium points, whose basins of attraction lie on
    both sides of the separatrix, for parameters $\delta_1=0.7, \beta_2=0.1,
    \delta_2=0.3, \tilde{\gamma}_1=2, \tilde{\gamma}_2=3, L_1=1, L_2=1$ and
    $K_2=2$. 
    Note that both $H$ and $S$ can persist if $H(0)$ and $S(0)$ lie on the
    upper right side of the separatrix even though they both cannot do so on
    their own. 
  }
  \label{fig:reduction21}
\end{figure}

This reduced system can show bistability, although not with positive fixed
points.
The origin is stable if $\beta_2<\delta_2$, but there can exist two more
interior fixed points, of which one is stable, as shown in  \Cref{fig:reduction21}. 
In this scenario, the symbiont is now also an obligate mutualist, which means both population sizes have to be sufficiently
large in order to allow their survival. 
The separatrix shown in  \Cref{fig:reduction21} discriminates the regions of
attractiveness of the zero and stable interior fixed point. 
Interestingly, low population sizes of one partner can be compensated by the
other, meaning that both populations can persist even when one partner's
population is small as long as the other is sufficiently large.

\subsection{A Reduction with Complex Interaction Terms}
\label{sec:reduction13}
The TFPV candidate 13
\begin{equation*}
  \tilde{\pi} = \left(\varepsilon \beta_2, \varepsilon \beta_3, \delta_1, \varepsilon
    \delta_2, \varepsilon \delta_3, \mu_1, \varepsilon \mu_2, \varepsilon \eta
  \right)
\end{equation*}
yields the affine variety
\begin{equation*}
  \cV(f^{(0)}) 
  = \cV\left(\mu_1 C \left(1 - \frac{H}{K_1}\right) - \delta_1 H \right)
  ,
\end{equation*}
which has dimension $s=2$.
We can choose the nonsingular point $x_0=0$. 
Then, the Jacobian of the fast part of the system is
\begin{equation*}
  Df^{(0)}(x_0) = 
\begin{pmatrix}
    - \delta_1 & 0 & \mu_1 \\
    0 & 0 & 0 \\
    0 & 0 & 0 
\end{pmatrix}
\end{equation*}
and~\eqref{eq:directsum} holds. 
Furthermore, the slow manifold is always attractive, since $\delta_1>0$. 
A product decomposition for $f^{(0)}$ can be chosen as
\begin{equation*}
  f^{(0)}(x) = P(x) \psi(x) = 
\begin{pmatrix}
    1 \\
    0 \\
    0 
\end{pmatrix}
\begin{pmatrix}
    \mu_1 C \left( 1 - \frac{H}{K_1} \right) - \delta_1 H
\end{pmatrix}
\end{equation*}
and $P$ and $\psi$ satisfy the conditions in \Cref{thm:reduction} 
\ref{thm:reduction-composition}.
The reduced system on the slow manifold is then given by
\begin{equation}
  \label{eq:reduction13}
  \begin{aligned}
    \dot{H} &= \left( \beta_3 - \delta_3 \right) H \left( 1 - \left( \frac{1}{K_1} +
      \frac{\beta_3 \delta_1}{\mu_1 \left( \beta_3 - \delta_3 \right) K_3 }\right) H
    \right) 
    + \frac{\eta \mu_1}{\delta_1} \left( 1 - \frac{H}{K_1} \right)^{2} S H \\
        \dot{S} &= \beta_2 S \left( 1 - \frac{S}{K_2} \right) - \delta_2 S - \eta S
        H + \mu_2 \left( 1 - \frac{S}{K_2}\right) \frac{\delta_1 H}{\mu_1 \left(
        1 - \frac{H}{K_1}\right)}
  \end{aligned}
\end{equation}
Note that the RHS of this reduction has a singularity at $H = K_1$. 
However, the system on the domain of interest $\mathopen[0,K_1\mathclose[
\times \mathopen[0,K_2\mathclose[$ is well-defined, since $H$ never
reaches $K_1$ due to
\begin{equation*}
  \dot{H}\vert_{H=K_1} = - \frac{\beta_3 \delta_1 K_1^2}{\mu_1 K_3} < 0.
\end{equation*}
The autarkic population $H$ grows logistically with base growth rate $\rho_3
\defl \beta_3-\delta_3$, which we assume to be positive. 
We define the rescaled capacity for $H$ as
\begin{equation*}
  \tilde{L}_1 \defl 
  \frac{\mu_1 \rho_3 K_3}{\beta_3 \delta_1 K_1 + \mu_1 \rho_3 K_3} K_1 
\end{equation*}
and introduce the parameter $\alpha \defl \delta_1/\mu_1$.
Note that $\beta_3>\delta_3$ implies $0<\tilde{L}_1<K_1$. 
Using these definitions, System~\eqref{eq:reduction13} can be simplified as
\begin{equation}
  \label{eq:reduction13_min}
\begin{aligned}
    \dot{H} &= \rho_3 H \left( 1 -  \frac{H}{\tilde{L}_1} \right) 
    + \frac{\eta }{\alpha} \left( 1 - \frac{H}{K_1} \right)^{2} S H \\
    \dot{S} &= \beta_2 S \left( 1 - \frac{S}{K_2} \right) - \delta_2 S 
    - \eta S H 
    + \alpha \mu_2 \left( 1 - \frac{S}{K_2}\right) \frac{H}{1 - \frac{H}{K_1}}
\end{aligned}
\end{equation}
 \Cref{fig:reduction13} shows the basic relation between the variants of the
logistic growth term for $H$. 
We can see that all other functions are dominated by
$g(H)\defl H(1-H/K_1)^{-1}$ due to the singularity.
Thus, the benefit for the photobiont becomes very large if $H$ is large or $S$
is small.
On the slow manifold we have $C(H) = \alpha g(H)$, which is monotonically
increasing for $0 \leq H < K_1$.  
Therefore, the lichen population only depends on the abundance of mycobionts and
becomes very large when $H$ is close to its capacity. 

\begin{figure}[tb]
  \begin{center}
    \begin{tikzpicture}
      \begin{axis}[
        xlabel=$H$,
        width=0.65\textwidth, height=0.4\textwidth,
        legend entries={
          $1-H/\tilde{L}_1$,
          $1-H/K_1$,
          $(1-H/K_1)^2$,
          $(1-H/K_1)^{-1}$
        },
        legend cell align = {left},
        legend style={
          at={(1.04,1)},
          anchor=north west,
          draw=none
        },
        grid=major,
        axis lines=left,
        xmin=0,
        enlargelimits=true,
        axis background/.style={fill=gray!05},
        xtick={0, 0.25, 1},
        xticklabels={$0$, $\tilde{L}_1$, $K_1$},
        ymin = -0.05,
        restrict y to domain = -1.0:2.0,
        domain=0:1,
        samples=200
        ]
        \addplot [orange!90!brown, semithick, smooth] {(1-4*x)};
        \addplot [teal!70!green, semithick, smooth] {(1-x)};
        \addplot [cyan!40!blue, semithick, smooth] {(1-x)^2};
        \addplot [purple, semithick, smooth] {1/(1-x)};
      \end{axis}
    \end{tikzpicture}
    \caption{Per-capita interaction terms in System~\eqref{eq:reduction13_min}
      for parameters $\beta_3 = 2$, $\delta_1 = 3$, $\delta_3 = 1$, $\mu_1 = 2$,
    $K_1 = 1$ and thus $\tilde{L}_1= 1/4$.}
    \label{fig:reduction13}
  \end{center}
\end{figure}

Evaluating the mutualism criterion for System~\eqref{eq:reduction13_min} yields
the benefit functions
\begin{equation*}
  b_1(S,H) = \frac{\eta S H (K_1-H)^2}{\alpha K_1^2}
\end{equation*}
and
\begin{equation*}
  b_2(S,H) = \frac{\alpha \mu_2 K_1 H(K_2-S)}{K_2(K_1 - H)} - \eta S H,
\end{equation*}
from which we can see that the net effect of $S$ on $H$ is always nonnegative.
The effect of $H$ on $S$ is not so clear. 
For $H<K_1$ we find $b_2(S,H)>0$ if and only if
\begin{equation*}
  S < \frac{\alpha \mu_2 K_1}{\alpha \mu_2 K_1 + \eta K_2 (K_1 - H)} K_2.
\end{equation*}
Thus, the photobiont's autarkic population only has an advantage if $S$ is
small or $H$ is large.
However, when estimating the overall effect of one partner on the other, we have
to consider the total populations, i.e.\ all individuals regardless of their
state (mutualistic or autarkic).

In contrast to the reductions discussed before, the redundant component, i.e.\
the one defined according to the slow manifold by the two remaining ones, is
not constant for this reduction.
Instead, \eqref{eq:reduction} yields
\begin{equation*}
  \dot{C} = \eta S H 
  + \alpha \rho_3 \frac{H}{1 - \frac{H}{K_1}}
  \left( 1 - \frac{\alpha \beta_3}{ \rho_3  K_3} \frac{H}{1 - \frac{H}{K_1} }
  \right). 
\end{equation*}
With this, the system with respect to total population sizes $X=X(H)=H+C(H)$ and
$Y=Y(S,H)=S+C(H)$ according to the slow manifold is given by
\begin{equation}
  \label{eq:reduction13_total}
  \begin{aligned}
    \dot{X} ={}& \rho_3 H \left( 1 -  \frac{H}{\tilde{L}_1} \right) 
    + \frac{\eta}{\alpha} \left(\alpha + \left( 1 - \frac{H}{K_1} \right)^{2}\right) S H 
    \\
               & + \alpha \rho_3 \frac{H}{1 - \frac{H}{K_1}}
               \left( 1 - \frac{\alpha \beta_3}{ \rho_3  K_3} \frac{H}{1 - \frac{H}{K_1} } \right)\\
    \dot{Y} ={}& \beta_2 S \left( 1 - \frac{S}{K_2} \right) - \delta_2 S 
    \\
               & + \alpha \frac{H}{1 - \frac{H}{K_1}}
               \left(\mu_2\left( 1 - \frac{S}{K_2}\right) + \rho_3 \left( 1 - \frac{\alpha \beta_3}{ \rho_3  K_3} \frac{H}{1 - \frac{H}{K_1} } \right) \right)
  \end{aligned}
\end{equation}
where
\begin{equation*}
  H = H(X) = \frac{1}{2} \left(K_1(1 + \alpha) + X 
  - 
  \sqrt{\left( K_1 (1 + \alpha) - X \right)^{2} + 4 \alpha K_1 X }\right)
\end{equation*}
and $S = S(X,Y)= Y - C(H(X))$. 
The derivation and a discussion of this can be found in Appendix
\ref{apx:derivationsystem13}.

Unfortunately this system is no longer rational and far from being easily
understood. 
We could already consider the full System~\eqref{eq:lichenmodel} for the total
populations $X$ and $Y$. 
By doing so, we get a rational reduction for the same TFPV, but only in the
components $Y$ and $C$. 
Therefore, even if this reduced system is much more elegant, we cannot
circumvent having to handle expressions involving a square root, because we need
the system to be defined for $X$ and $Y$ to discuss the effect of the
interaction between the two partners.

Nevertheless, the closed form of the system with respect to total population
sizes allows us to obtain some information about the reduction.
Firstly, we find that $H(X)$ increases monotonically with $X$ and
$\lim_{X\to\infty} H(X) = K_1$. 
The ratio of mycobionts in autarkic state compared to their total abundance is
\begin{equation*}
  \frac{H}{X} = \frac{H}{H + C(H)} = \frac{K_1 - H}{K_1(1 + \alpha) - H} \xrightarrow[H\to K_1]{} 0
\end{equation*}
and decreases monotonically with $H$.
Thus, if the autarkic population is close to its capacity, almost all
mycobionts are found in lichens.
Similarly, for the symbiont, we find
\begin{equation*}
  \frac{S}{Y} = \frac{S}{S + C(H)} = \frac{(K_1-H)S}{ (K_1-H)S + \alpha K_1 H}.
\end{equation*}
Most importantly, we can compute the benefit functions for
System~\eqref{eq:reduction13_total} in order to evaluate the overall net
effects.
This yields
\begin{equation*}
\begin{aligned}
    b_1(X,Y) ={}& \eta H(X) Y \left( 1 + \frac{1}{\alpha} 
      \left( 1 -\frac{H(X)}{K_1} \right)^{2}\right) \\
    b_2(X,Y) ={}& \frac{\alpha K_1 H(X)}{\left( K_1 - H(X) \right) K_2}
      \Biggm(
        \frac{\alpha K_1 H(X) \left(\mu_2 K_3 - \beta_3 K_2 - \beta_2 K_3 \right)
        }{\left(K_1 - H(X) \right) K_3} \\
&        - K_2 \left( \beta_2 - \delta_2 - \mu_2 - \rho_3 \right) 
        + Y \left( 2 \beta_2 - \mu_2\right) 
    \Biggm)
\end{aligned}
\end{equation*}
Thus, as we have seen before, the mycobiont always has a net benefit, whereas
the benefit for the symbiont depends on the parameters and the population sizes
of the two partners.

\subsection{A Reduction with Source and Sink Dynamics}
\label{sec:reduction18}
The TFPV candidate 18
\begin{equation*}
  \tilde{\pi} = \left(\varepsilon \beta_2, \beta_3, \varepsilon \delta_1,
    \varepsilon \delta_2, \delta_3, \varepsilon \mu_1, \varepsilon \mu_2,
    \varepsilon \eta \right)
\end{equation*}
admits the affine variety
\begin{equation*}
  \cV(f^{(0)})
  = \cV(C) \cup \cV\left(\beta_3 \left(1-\frac{C}{K_3}\right) - \delta_3\right)
  \defr Y_1 \cup Y_2
\end{equation*}
with two irreducible components of dimension $s=2$ 
(assuming $\beta_3 \neq \delta_3$). 
Here, we will only consider the reduction onto $Y_2$.
We can choose the nonsingular point 
$x_0 = \left( H', S', \frac{\beta_3 -\delta_3}{\beta_3} K_3 \right)$ for
arbitrary $H',S'>0$. 
Then
\begin{equation*}
  Df^{(0)}(x_0) = 
\begin{pmatrix}
    0 & 0 & 0 \\
    0 & 0 & 0 \\
    0 & 0 &  - (\beta_3 - \delta_3)
\end{pmatrix}
\end{equation*}
satisfies~\eqref{eq:directsum}. 
Therefore, $\tilde{\pi}$ is indeed a TFPV and the slow manifold 
$Y_2$ is attractive if $\beta_3>\delta_3$. 
With the product decomposition
\begin{equation*}
  f^{(0)}(x) = P(x) \psi(x) = 
\begin{pmatrix}
  0 \\
  0 \\
  C 
\end{pmatrix}
\begin{pmatrix}
  \beta_3 ( 1 - \frac{C}{K_3}) - \delta_3 
\end{pmatrix}
\end{equation*}
and $C^\star \defl \frac{\beta_3 - \delta_3}{\beta_3} K_3$ the reduced system is given as
\begin{equation}
\begin{aligned}
    \label{eq:reduction19}
    \dot{H} &=  - \delta_1 H - \eta S H 
      + \mu_1 C^\star \left( 1 - \frac{H}{K_1}  \right) \\
    \dot{S} &= \beta_2 S \left( 1 - \frac{S}{K_2} \right) - \delta_2 S - \eta S H
      + \mu_2 C^\star \left( 1 - \frac{S}{K_2} \right) 
\end{aligned}
\end{equation}
In this reduction, the complex is always present and serves as a sink for the
autarkic populations if the latter are large, and as a source, if they are
small, which means the net benefit depends on the sizes of the autarkic
populations.
However, this cannot be seen with either \Cref{def:strongmutualismcriterion} or
\ref{def:mutualismcriterion}.
To obtain the system with respect to total population sizes, we have to
substitute $X=H-C^\star$ and $Y=S-C^\star$ (note that $\dot{C}=0$). 
This system is then defined on $D=\mathopen[{C^\star},{\infty}\mathclose[^2$ and
we see that the direct effect of the two partners is always nonpositive:
\begin{equation*}
  b_1(X,Y) = b_2(X,Y) = -\eta (X-C^\star)(Y-C^\star) = -\eta S H \leq 0.
\end{equation*}
Does this mean the mutualism criterion is of limited value?
One may argue that \Cref{def:mutualismcriterion} is not helpful here because the
interpretation of the original system is hidden away in the reduction and the
population size of the complex now effectively became a parameter.
More precisely, from System~\eqref{eq:reduction19} alone we cannot see that
births from the complex are only possible because $H$ and $S$ are present in the
first place.
The costs associated to the formation of lichens are described by the term
$-\eta S H$, but the benefit for both species in form of additional births from
the complex are not depending on their partner's populations, because the lichen
population is always at its carrying capacity.

This demonstrates that the mutualism criterion may fail to characterize
interactions according to our biological understanding if static effects are
prevailing. 
However, this also shows the strength of this modelling approach. 
In fact, the embedding of System~\eqref{eq:reduction19} as a special case of our
super model \eqref{eq:lichenmodel} and the resulting interpretation allow us to
circumvent this problem.
Here, we can compare the terms that correspond to the interaction directly.
Taking this into account, we can see that the interaction has a net benefit for
the photobiont if
\begin{equation*}
  \mu_2 C^\star \left( 1 - \frac{S}{K_2} \right) - \eta S H > 0 
  \iff 
  S < \frac{\mu_2 C^\star}{\eta K_2 H + \mu_2 C^\star} K_2 \leq K_2.
\end{equation*}
Similarly, the mycobiont gains an advantage, if
\begin{equation*}
  H < \frac{\mu_1 C^\star}{\eta K_1 S + \mu_1 C^\star} K_1 \leq K_1.
\end{equation*}
This means the advantage for each partner is maximal if both autarkic
populations are small. 
The reason for this behaviour is that the formation of $C$ still occurs with
rate $\eta S H$, but the lichen population has a fixed size.
If many autarkic individuals are present, they will quickly form new lichens.
Since the lichen population is always in its carrying capacity on the slow
manifold, any additional individuals will die immediately.
If on the other hand the autarkic populations are small, there will be more
births from the lichen complex than autarkic individuals lost due to the
formation of new lichens. 
Thus, in this scenario lichens form a source or sink for the autarkic
populations of photo- and mycobiont, depending on their respective population
sizes.  

This also explains another interesting aspect of System~\eqref{eq:reduction19}:
No population can go extinct. 
Instead, due to the fact that the complex is always present, each autarkic
population starts growing even when no individuals are present.
This behaviour is comparable to a system with migration, where extinct local
populations can be re-established via immigration from another patch. 

\subsection{Reductions for General TFPVs}
\label{sec:alternativereductions}
Now we want to consider reductions for TFPVs that are not slow--fast separations
of rates. 
These can be found using the method described in  \Cref{sec:findingtfpvgroebner}
and for System~\eqref{eq:lichenmodel} computation of the elimination ideal
$I_{\pi}$ is still feasible (see supplementary material).
In~\textcite{kruff2019}, the elimination ideal is generated by monomials, which
implies that every TFPV is a slow--fast separation of rates.
In our case, $I_{\pi}$ does not contain any monomials, which can be
checked by computing the saturation of $I_{\pi}$:
An ideal $J\subseteq K[x_1,\dots,x_n]$ contains a monomial if and only if
\begin{equation*}
  J : \langle x_1\cdots x_n \rangle^\infty 
  = K[x_1,\dots, x_n], 
\end{equation*}
which can be computed algorithmically~\cite{jensen2017}. 
In order to characterize TFPVs that are not slow--fast separations further, we
can compute a primary decomposition for $I_{\pi}$, which turned out to be not
feasible for our model.
Instead, we can consider cases for the vanishing of $I_{\pi}$ to find at least
some of these general TFPVs. 
However, all the corresponding reductions we found are special cases of
reductions corresponding to slow--fast separations, but we will illustrate the
general procedure with the following example.

In our case, the elimination ideal $I_{\pi}$ vanishes if $\phi \defl \mu_1 -
\mu_2 = 0$ together with at least $\beta_2,\delta_1,\delta_2,\eta=0$. 
Writing System~\eqref{eq:lichenmodel} as
\begin{equation*}
\begin{aligned}
    \dot{H} &= - \delta_1 H - \eta SH + (\phi + \mu_2) C \left(1-\frac{H}{K_1}\right) \\
    \dot{S} &= \beta_2 S \left(1-\frac{S}{K_2}\right) - \delta_2 S - \eta SH + \mu_2 C \left(1-\frac{S}{K_2}\right) \\
    \dot{C} &= \beta_3 C \left(1-\frac{C}{K_3}\right) - \delta_3 C + \eta SH
\end{aligned}
.
\end{equation*}
allows us to apply the routine for finding slow--fast separations that are TFPVs. 
This yields, among others, the candidate
\begin{equation*}
  \tilde{\pi} = (\varepsilon\beta_2, \beta_3, \varepsilon\delta_1,
  \varepsilon\delta_2, \delta_3, \varepsilon\phi, \mu_2, \varepsilon\eta).
\end{equation*}
The slow manifold is $\cV(f^{(0)}) = \set{ (H,S,0) \mid H,S \in \R }$ and
the reduction is exactly System~\eqref{eq:reduction21_rewrite} with
$\mu_1=\mu_2$, as suggested by $\phi$ being a small parameter together with
$\beta_2,\delta_1,\delta_2$ and $\eta$. 

\section{Discussion}
\label{sec:discussion}
Singular perturbation theory provides useful methods for modelling dynamical
systems.
Its core idea is the occurrence of a small parameter $\varepsilon>0$ that
perturbs the original system and separates its components into slow and fast
ones, such that the system evolves on two different time scales.
Intuitively, the fast components evolve so quickly, that the slow ones hardly
see any change. 
Taking the limit $\varepsilon\to0$, we therefore approximate the fast part with
its steady state (if it exists).
The precise formulation of this approach goes back to~\textcite{tikhonov1952} and
can be found in \cite{verhulst2007} in its present-day form.
A coordinate-free approach to singular perturbation theory was pioneered
by~\textcite{fenichel1979} and a recent overview of the theory can be found
in~\cite{wechselberger2020}.

In terms of population dynamics, singular perturbation methods aid the
mathematical analysis of models. 
They can also be used to derive and justify new conceptual models from
carefully crafted super models.
Roughly speaking, we consider such a super model to be a mathematical system
that depicts the relevant (ecological) aspects of the real world system with
sufficient detail.
This typically leads to high dimensional dynamical systems that are difficult to
analyse mathematically and hence offer little insight. 
But these models can potentially relate closer to the real world system, as
the focal actors and processes may be considered explicitly~\cite{metz2005}.
This often makes it possible to derive the equations from first principles and
enhances their interpretability. 
Model reduction via time scale separation then allow us to obtain lower
dimensional and potentially much simpler models that inherit properties of the
super model, most importantly their biological justification and interpretation.
In other words, singular perturbation methods allow us to translate biology into
mathematics in detail, without compromising on the feasibility of model
analysis.
This effectively mitigates the trade-off between model complexity and realism.

However useful singular perturbation theory is, the traditional approach of
ad hoc model reduction with Tikhonov's theorem has several shortcomings.
Most importantly, its application is only possible for a system in Tikhonov
normal form, i.e.\ if the components of the ODE system can be separated into
slow and fast ones.
In population dynamics this usually means some population density or resource
concentration is considered to be in quasi-steady state (as in e.g.\
\cite{ludwig1978,fishman2010,revilla2015}, see~\cite{abbott2020} for an
overview).
However, identifying which components evolve slowly is typically not
straightforward. 
In many situations we want to discriminate between slow and fast processes
instead of components, which means the quasi-steady state assumption is simply
too restrictive. 

For ad hoc reductions, one may introduce artefacts from manual intervention and
additional approximations that would not arise from the application of
Tikhonov's theorem alone. We might be tempted to think of expressions as being
small compared to some particular others, but since computing the reduction
involves taking a limit, they are actually small compared to \emph{all} others.

The algebraic approach of Tikhonov--Fenichel reductions introduced
by~\textcite{goeke2012} (see also~\cite{goeke2013a,goeke2014,goeke2015}) used
in this paper overcomes the problems stated above. 
In particular, we are able to obtain model reductions arising from time scale
separations of processes instead of components, which allows for a more natural
and much finer distinction of (biological) scenarios.
This is not possible for quasi-steady state reductions directly, as it requires
a coordinate transformation into Tikhonov normal form~\cite{noethen2011}. 

The main feature of Tikhonov--Fenichel reductions utilized in this paper is that we
do not have to consider each scenario on its own. 
By evaluating algebraic conditions for the existence of a formal reduction for a
particular super model, we are able to find all possible slow--fast separations
of rates admitting a reduction in the sense of Tikhonov and Fenichel entirely
algorithmically~\cite{goeke2015}.
Computing the corresponding reduced systems is then done quasi-automatically,
which eliminates the need to decide a priori what rates or components should be
considered small and reduces the risk of introducing errors. 
Furthermore, since sufficient conditions are known (see
\Cref{thm:tfpvsufficient}), we do not have to check manually that the conditions
in Tikhonov's theorem are satisfied.

Our main contribution to Tikhonov--Fenichel reductions is the development of the
free and open source Julia package \texttt{Tikhonov\-Fenichel\-Reductions.jl}
\cite{apelt2024}, which allows users to conveniently apply the algebraic
approach to time scale separations for polynomial ODE systems in a
straightforward manner.
The essential functionality provided by our package is an implementation of
algorithms to find all critical parameters that yield a formal reduction and the
convenient computation of the corresponding reduced systems.

Using Tikhonov--Fenichel reductions, we are thus able to compute reduced systems
systematically. 
Naturally, we then want to analyse and interpret the resulting systems. 
In our case, the most fundamental question is whether the system is still
mutualistic.
We think that the mutualism criterion in \Cref{def:mutualismcriterion} should be
applied to decide under which circumstances a two-species model represents
mutualism, since it measures the absolute effect of the presence of one
population on the other instead of the trend of the effect --- as is the case for
the strong mutualism criterion that is typically used (e.g.\
in~\cite{brauer1985,neuhauser2004,wang2011}) and which corresponds to the
definition of a cooperative system~\cite{smith1995} in our setting.
Moreover, the mathematical formulation directly reflects what is measured in
experiments concerning the same question~\cite{bronstein1994}.
The seemingly subtle distinction between the two criteria proved to be very
important, as they may indeed classify the same interaction differently
(see  \Cref{fig:red12_contour}).

We furthermore observed that we have to be careful in examining the net benefit
of the interaction for the reductions of our super model \eqref{eq:lichenmodel}. 
As we have seen, it is important to consider its underlying structure.
In particular, we need to apply the criterion to the reduced systems written in
terms of total population sizes to estimate the overall effect and obtain results
comparable to other models. 
We have observed the general pattern that the host, an ecologically obligate
mutualist, always tends to have a positive net benefit, whilst the effect of the
interaction depends on the parameters and population sizes of both species for
the facultatively mutualistic symbiont and may even become negative.
When the mutualistic birth rate for the symbiont is a small parameter, but not
the one for the host, the mutualistic relation may shift towards parasitism
entirely, as we have seen in  \Cref{sec:reduction20}.

Besides the breakdown of mutualism, Tikhonov--Fenichel reductions have revealed
several scenarios that were not obvious from the super model
\eqref{eq:lichenmodel} alone.
We have seen in  \Cref{sec:reduction18} that the lichen population can be a
source or sink for the autarkic populations of host and symbiont if births of
the lichen complex exceed deaths and both processes occur much quicker compared
to all others.
However, this scenario seems to be rather unrealistic in terms of the biology of
lichens.

We found via numerical analysis that bistability can occur for the super model
(even with multiple stable interior fixed points) and its reductions, as can be
seen in  \Cref{fig:red12_bistability,fig:lichenmodelbistable} (see also
 \Cref{fig:reduction21}).
It remains unclear, whether this is a merely mathematical effect or if this
behaviour can be observed in real world systems.

Functional responses are essentially implicit descriptions of the effects of the
interaction between two populations. 
Originally they were used in the context of predator--prey systems, but have been
generalized to other types of interactions~\cite{holland2002}.
In our super model \eqref{eq:lichenmodel}, we considered the interaction between
the potentially mutualistic partners explicitly by introducing the complex $C$
formed by the individuals that are actually interacting.
As a consequence, we do not rely on a particular choice of a functional response. 
This allows us to consider the mechanisms behind the effects of one population
on the other directly instead of assuming a certain saturation of the benefit.
Moreover, the mathematical formulation of the model closely follows the
definition of mutualism, i.e.\ the increase in fitness due to the interaction is
modelled by additional births.
Performing Tikhonov--Fenichel reductions then yields ODE systems that resemble
the conceptual models typically used to describe population dynamics for two
species interacting mutualistically (see e.g.\ \cite{hale2021} for an
overview).

The important difference is that this method provides a good interpretation and
justification for the particular choice of the functional responses in different
scenarios.
This is especially important as the classical approach to functional responses
relates the same mathematical expression (up to a multiplicative constant) to
two completely different processes. 
For instance in the famous model by~\textcite{rosenzweig1963}, the type II
functional response describes the process of predation and births --- two
processes that are fundamentally very different and arguably occur on two
different time scales.
We believe that this correspondence should therefore be very carefully
justified, which in case of the Rosenzweig--MacArthur model can be done with
Tikhonov--Fenichel reductions, as demonstrated by~\textcite{kruff2019}.

The reductions for our super model \eqref{eq:lichenmodel} revealed several
different interaction terms.
Besides the typical type II functional response that occurs in System
\eqref{eq:reduction12}, we have found a type I functional response with a logistic
conversion rate in Systems \eqref{eq:reduction20} and
\eqref{eq:reduction21_rewrite}. 
We can interpret this as additional births resulting from the interaction with
successful establishment of the offspring being governed by intraspecific
density dependence.
The most unconventional interaction terms occur in System
\eqref{eq:reduction13_min}, 
which approximates our super model \eqref{eq:lichenmodel} when deaths of the
autarkic host and births from the lichen complex into its autarkic population
occur quickly compared to the other processes.

Alongside its practical advantages, Tikhonov--Fenichel reductions allow us to
gain ecological insight via the interpretation of the reduced systems in terms
of the biologically detailed super model.
Firstly, the separation of rates into slow and fast directly tells us on which
time scales the corresponding processes evolve.
Secondly, the slow manifold describes exactly how the components that were
reduced behave.
Thus, even though the dynamics of the original system can be explained by the
reduction (in the particular scenario), the super model and its biological
detail still yield information that is not present in the reduced system alone.
Furthermore, in many cases there are multiple reductions onto the same slow
manifold, which is not easily found with the traditional approach.
And finally, the resulting implicit descriptions of aspects of the system can
be traced back to their explicit formulation in the super model.
In our case, the resulting functional responses represent a simplified
description of the mutualistic interaction, which is defined explicitly in
System~\eqref{eq:lichenmodel}.

In conclusion, we strongly agree with the idea put forward by~\textcite{metz2005},
that \enquote{oversimplified models are good tools for discovering phenomena.
But their eventual justification should come from their embedding in a larger
class of models, some members of which connect more directly to the real
biological world.}
We believe that the approach presented in this paper is one possibility
to achieve this in a mathematically sound way.

\section*{CRediT Authorship Contribution Statement}
\textbf{Johannes Apelt:} Conceptualization, Formal Analysis, Methodology,
Software, Writing;
\textbf{Volkmar Liebscher:} Conceptualization, Formal Analysis, Methodology,
Supervision, Writing.

\section*{Funding}
This work was supported by a scholarship awarded by the University of Greifswald
according to the \enquote{Landesgraduiertenförderungsgesetz (LGFG) MV}.

\section*{Acknowledgements}
We thank Sebastian Walcher and Alexandra Goeke for the development of
Tikhonov--Fenichel reduction theory and for making us aware of its many
advantages, 
Leonard Schmitz for the helpful discussions about aspects of the computational
algebra used in this paper,
Frank Hilker for his interest in our work and bringing references regarding the
phenomenon of bistability to our attention, 
and Hans Metz for his ideas and the fruitful discussion about ecological
modelling.
We thank the editor Sebastian Schreiber and two anonymous reviewers for their
helpful and detailed remarks and for bringing related literature to our
attention.

\appendix
\section{Mathematical Preliminaries}
\label{apx:mathematicalpreliminaries}
Here we list the basic definitions and facts from algebraic geometry and
commutative algebra used in this paper.
We will mostly follow~\textcite{cox2015} and~\textcite{kunz2013}. 

We denote a polynomial ring over a field $K$ with indeterminates $x_1,\dots,x_n$
as $K[x_1,\dots,x_n]$ or $K[x]$ in short.
$K(x)$ denotes the set of rational functions over $K$, i.e.\ the field of
fractions of $K[x]$.
An ideal is a subset of a ring that contains the zero element and is closed under
addition as well as multiplication by arbitrary ring elements.
The radical of an ideal $I\subseteq K[x]$ is
\begin{equation*}
  \sqrt{I} \defl \set{f\in K[x] \mid \exists k\in\N_{>0}: f^k \in I}.
\end{equation*}
For a finite subset $F\subseteq K[x]$ or $F=(f_1,\dots,f_m)\in K[x]^m$, the
ideal generated by the polynomials in $F$ is
\begin{equation*}
  \langle F \rangle \defl \setlr{\sum_{i=1}^m h_i f_i \midvert h_1,\dots,h_m \in K[x]}.
\end{equation*}
Every ideal in a polynomial ring is finitely generated (this is known as the
Hilbert Basis Theorem).
A special type of generating set for an ideal $I\subseteq K[x]$ is a Gr\"obner
basis. 
Given any generating set for $I$ and a monomial ordering, one can compute a
Gr\"obner basis using Buchberger's algorithm. 
A thorough discussion of Gr\"obner bases and monomial orderings is beyond the
scope of this paper, interested readers can refer to~\cite{cox2015}.
One characterizing property of Gr\"obner bases is the uniqueness of the remainder
upon division of a polynomial by a Gr\"obner bases, which is not the case for all
sets of polynomials. 
This yields the following definition.
The normal form of a polynomial $p\in K[x]$ with respect to an ideal $I$,
denoted $\text{NF}(p,I)$, is the remainder upon division of $p$ by a Gr\"obner
basis of $I$. 

For a field $L$ with subfield $K\subseteq L$ and $F\subseteq K[x]$ we define the
affine $K$-variety in the affine space $L^n$ as
\begin{equation*}
  \cV_{L}(F) \defl \set{x \in L^n \mid \forall f \in F: f(x)=0}.
\end{equation*}
Conversely, any set $V\subseteq L^n$ for which there exists
a finite set $F\subseteq K[x]$ such that $V=\cV_L(F)$ is called an
affine $K$-variety. 
Whenever we omit the subscript, it is implied that the field of definition $K$
and the coordinate field $L$ are equal. 
Note that we always have $\cV_L(F) = \cV_L(\langle F \rangle)$.
An affine variety $V$ is said to be irreducible if $V=V_1 \cup V_1$, for affine
varieties $V_1$ and $V_2$, implies $V=V_1$ or $V=V_2$. 

An important property of an affine variety is its dimension, which can be
defined in different ways. 
Firstly, we can make use of the Zariski topology, which yields a definition that
relates closely to our intuitive geometric understanding of dimension. 
A set $X \subseteq L^n$ is closed in the Zariski topology if it is an affine
variety and open if it is the complement of a closed set. 
Let $X$ be a topological space.
A subset $Y \subseteq X$ is called irreducible if it cannot be written as a
proper union $Y=Y_1\cup Y_2$ of closed subsets $Y_1,Y_2\subseteq X$. 
The topological dimension of $X$ is
\begin{equation*}\dim X \defl \sup
  \set{k\in\N \mid Z_0 \subset Z_1 \subset \dots \subset Z_k \text{ dist.\ irr.\
  cl.\ subsets of } X},
\end{equation*} 
where $\emptyset \neq Z_0 \subset Z_1 \subset \dots \subset
Z_k\subseteq X$ is an ascending chain of distinct irreducible closed subsets
of $X$.
The dimension of an affine variety is its dimension as a topological space with
the Zariski topology.

Another definition makes use of the Krull dimension, which is a purely algebraic
concept. 
Let $V\subseteq L^n$ be an affine variety. The vanishing ideal of $V$ is
defined as
\begin{equation*}
  \cI(V) \defl \set{f \in K[x] \mid \forall a \in V : f(a) = 0 }
\end{equation*}
and its coordinate ring is
\begin{equation*} K[V] \defl K[x]/\cI(V) = \set{[f] \mid f\in K[x]},
\end{equation*} 
where $[f] \defl \set{g\in K[x] \mid f-g \in \cI(V)}$.

Let $R$ be a ring. 
An ideal $\mathfrak{p}\in R$ is prime if $fg\in\mathfrak{p}$ implies
$f\in\mathfrak{p}$ or $g\in\mathfrak{p}$. 
The set of all primes $\mathfrak{p}\neq R$ in $R$ is denoted $\Spec(R)$ and the
height of a prime ideal is
\begin{equation*} 
  h(\mathfrak{p}) \defl 
  \sup \set{k\in\N \mid \mathfrak{p}_0 \subset \mathfrak{p}_1 \subset \dots
    \subset \mathfrak{p}_k = \mathfrak{p} : \mathfrak{p}_i \in \Spec(R),
    \mathfrak{p}_i\neq\mathfrak{p}_{i+1}}.
\end{equation*}
The Krull dimension of $R$ is then defined as
\begin{equation*}
  \dim R \defl \sup \set{h(\mathfrak{p}) \mid \mathfrak{p}\in\Spec(R)}.
\end{equation*}
The Krull dimension of an affine variety is defined as the Krull dimension of
its coordinate ring.
Similarly, the Krull dimension of an ideal $I$ is the dimension of the ring
$K[x]/I$.
Note that $\dim I = \dim \sqrt{I}$, because $I$ and $\sqrt{I}$ are contained
in exactly the same prime ideals and there is an inclusion preserving one-to-one
correspondence between the ideals containing $I$ in $K[x]$ and the ideals in
$K[x]/I$.

Whenever the coordinate field of the underlying affine space is algebraically
closed, the two notions of dimension coincide, 
otherwise the topological dimension might be smaller.
For us this is rather unfortunate, because we can easily compute the Krull
dimension using computer algebra software, but for our application we are
interested in the topological dimension of real affine varieties.
However, the following concepts help circumvent this problem in our setting.

The linear part of a polynomial $f\in K[x]$ at $a$ is defined as
\begin{equation*}
  d_a(f) = \frac{df}{dx_1}(a)(x_1-a_1) +\cdots+ \frac{df}{dx_n}(a)(x_n-a_n)
\end{equation*}
and the tangent space at a point $a$ of an affine variety $V$ is
\begin{equation*}
  T_a(V) = \cV\left(\set{d_a(f) \mid f\in \cI(V)}\right),
\end{equation*}
which is a translate of a linear subspace.
A point $x_0$ of an affine variety is nonsingular if the dimension of
the tangent space equals the topological dimension of the variety at $x_0$. 

Since we will only be concerned with affine $\R$-varieties in $\R^n$ containing
a nonsingular point, 
the two notions of dimension above do actually coincide~\cite{marshall2008}
(as cited in~\cite[][Thm.\ 2.4.]{harris2023}). 
Thus, we can check whether an affine variety $\cV_{\C}(I)$ for an ideal
$I\subseteq\R[x]$ contains a real nonsingular point $x_0$.

If the irreducible affine variety $V=\cV_{\C}(f_1,\dots,f_m)$ has dimension
$s$ and for $a\in V$ the Jacobian of $f=(f_1,\dots,f_m)$ satisfies $\text{rank}
\; Df(a)  = n - s$, then $a$ is a nonsingular point of $V$.
A point on $V$ is nonsingular if and only if this equality holds for
$f_1,\dots,f_m$ generating $\cI(\cV_\C(I))$. 
Note that this criterion is closely related to the implicit function theorem.

Hilbert's Nullstellensatz tells us that $\cI(\cV_\C(I)) =
\sqrt{I}$.
An ideal $I$ is primary if $fg\in I $ implies $f\in I$ or 
$g\in\sqrt{I}$.
Every ideal $I\subseteq K[x]$ has a minimal primary decomposition, i.e.\ there exist
primary ideals $Q_i$, such that $\bigcap_{i=1}^m Q_i = I$, the $\sqrt{Q_i}$
are distinct and $Q_i \not\supseteq \bigcap_{i\neq j} Q_j$.
Such a minimal decomposition can be computed algorithmically~\cite{gianni1988}
and also decomposes the corresponding affine variety:
$\cV_{\C}(I) 
= \bigcup_{i=1}^m \cV_{\C}(Q_i) 
= \bigcup_{i=1}^m \cV_{\C}(\sqrt{Q_i})$. 
Because each $\sqrt{Q_i}$ is a prime ideal, the corresponding varieties are the
irreducible components of $\cV_{\C}(I)$.
The dimension of $\cV_{\C}(Q_i)$ can be computed as the Krull dimension
of $Q_i$, since $\dim \sqrt{Q_i} = \dim Q_i$.
Now, if $\cV_{\C}(Q_i)$ contains a real nonsingular point $x_0$,
the topological dimension of its real part, i.e
$\cV_{\R}(Q_i)\ni x_0$, is the same as $\dim Q_i$. 

\section{Proofs and Computations}
\label{apx:proofsandcomputations}

\subsection{Proof of  \Cref{thm:solutionbounded}}
\begin{proof}
  Let $x(t)=(H,S,C)(t)$ and System~\eqref{eq:lichenmodel} be written as
  $\dot{x}=f(x)$. 
  Solutions with initial value $x(0)\in\R^3$ exist and are unique.
  We use Nagumo's theorem (1942)~\cite[Theorem 3.1]{blanchini1999} to show
  that
  the closed and convex set $D$ is positively invariant under
  \eqref{eq:lichenmodel}. 
  Let $\mathcal{C}_D(x)$ be the tangent cone of $D$ at $x$, i.e.\
  \begin{equation*}
    \mathcal{C}_D(x) = \setlr{z\in\R^3 \midvert \liminf_{h\to0} \frac{\text{dist}(x + hz, D)}{h} = 0}.
  \end{equation*}
  We need to show that the vector field defined by $f$ points inwards or is
  tangential to $D$ along its boundary $\partial D$, i.e.\ $\forall x\in\partial
  D : f(x)\in\mathcal{C}_D(x)$.
  We consider points on the faces of the cuboid $D$ first.

  Let $H\in[0, K_1], S\in[0, K_2]$ and $C\in[0, \tilde{K}_3]$ be arbitrary. 
  For the lower bounds we find
  \begin{equation*}
    \begin{aligned}
      f_1((0,S,C)) &= \mu_1 C \geq 0 \\
      f_2((H,0,C)) &= \mu_2 C \geq 0 \\
      f_3((H,S,0)) &= \eta S H \geq 0
    \end{aligned}
  \end{equation*}
  and for the upper bounds
  \begin{equation*}
    \begin{aligned}
      f_1((K_1,S,C)) &= -\delta_1 K_1 - \eta K_1 S < 0 \\
      f_2((H,K_2,C)) &= -\delta_2 K_2 - \eta K_2 H < 0 \\
      f_3((H,S,\tilde{K}_3)) &= \eta H S - 2 \eta K_1 K_2 \leq - \eta K_1 K_2 < 0
    \end{aligned}
  \end{equation*}
  Thus, $f(x)\in\mathcal{C}_D(x)$ for any point $x$ in the interior of the faces
  of $D$.

  The tangent cone along the edges and vertices is the intersection of the
  tangent cones of the adjacent faces.
  For $x$ lying on an edge or being a vertex of $D$, $f(x)$ is a combination of
  the corresponding directions $f_i(x)$ as above, which means the vector field
  points inwards or is tangential to $D$ for all $x\in\partial D$. 
  Therefore, $D$ is invariant under \eqref{eq:lichenmodel} in light of Nagumo's
  theorem.
\end{proof}

\subsection{Proof of \Cref{thm:existenceinteriorfp}}
\begin{proof}
  Let $x_0^\star=(0,0,0)$ and
  $x_1^\star=\left(0,\frac{\beta_2-\delta_2}{\beta_2} K_2,0\right)$. 
  We show that these two points are the only fixed points in $\R_{\geq 0}^3$
  where some components are equal to zero. 
  First, note that $H=0$ or $S=0$ implies $C=0$. 
  Conversely, $C=0$ implies $H=0$, since $\dot{H}\vert_{C=0} = H(-\delta_1 -
  \eta S)$ vanishes if $H=0$ or $S=-\delta_1/\eta<0$.
  Similarly, 
  \begin{equation*}
    \dot{S}\vert_{H,C=0}=\beta_2 S \left(1-\frac{S}{K_2}\right) - \delta_2 S = 0
    \iff 
    S=0 \text{ or } S=\frac{\beta_2-\delta_2}{\beta_2} K_2.
  \end{equation*}
  Thus, the only fixed points with some components equal to 0 are $x_0^\star$
  and $x_1^\star$.

  For the second part we assume $H,S,C > 0$ s.t.\ the RHS of
  \eqref{eq:lichenmodel} vanishes. 
  Then, we can rewrite the last equation in \eqref{eq:lichenmodel} as 
  \begin{equation*} -\eta S H = \left( \beta_3\left( 1-\frac{C}{K_3} \right) - \delta_3 \right)C. 
  \end{equation*}
  Substituting this in the first equation yields
  \numberwithin{equation}{section}
  \setcounter{equation}{0}
  \begin{equation}
    \label{eq:H(C)}
    H(C) = \frac{\left( \beta_3\left( 1-\frac{C}{K_3} \right) + \mu_1 - \delta_3 \right)C}{\frac{\mu_1}{K_1}C + \delta_1}
  \end{equation}
  and we can substitute this back into the last equation to get
  \begin{equation}
    \label{eq:S(C)}
    S(C) = \frac{1}{\eta}\frac{\left( \delta_3 - \beta_3\left( 1-\frac{C}{K_3} \right) \right) \left( \frac{\mu_1}{K_1}C + \delta_1 \right) }{\beta_3\left( 1-\frac{C}{K_3} \right) + \mu_1 - \delta_3} .
  \end{equation}
  We can then use \eqref{eq:H(C)} and \eqref{eq:S(C)} to write the second equation as
  \begin{align*}
    0 ={}& 
    \left(\beta_3\left(1 - \frac{C}{K_3}\right) + \mu_2 - \delta_3\right)C
    + \left(\beta_2\left(1 - \frac{S}{K_2}\right) - \delta_2 - \frac{\mu_2}{K_2}C\right)S \\
    ={}& \left(\beta_3\left(1 - \frac{C}{K_3}\right) + \mu_2 - \delta_3\right)C \\ 
       & \quad + \left(\beta_2 - \delta_2 - \frac{\beta_2}{K_2} \left( \frac{1}{\eta}\frac{\left( \delta_3 - \beta_3\left( 1-\frac{C}{K_3} \right) \right) \left( \frac{\mu_1}{K_1}C + \delta_1 \right) }{\beta_3\left( 1-\frac{C}{K_3} \right) + \mu_1 - \delta_3} \right) - \frac{\mu_2}{K_2}C\right) \\
       &\quad \cdot
       \left( \frac{1}{\eta}\frac{\left( \delta_3 - \beta_3\left( 1-\frac{C}{K_3} \right) \right) \left( \frac{\mu_1}{K_1}C + \delta_1 \right) }{\beta_3\left( 1-\frac{C}{K_3} \right) + \mu_1 - \delta_3} \right)
  \end{align*}
  The zero loci of this rational function are exactly the roots of the fourth
  order polynomial
  \begin{equation}
    \label{eq:polynomC}
    \begin{split}
      p(C) \defl{}& \eta^2\left( \beta_3\left( 1-\frac{C}{K_3} \right) + \mu_1 - \delta_3 \right)^2
      \left(\beta_3\left( 1-\frac{C}{K_3} \right) + \mu_2 - \delta_3\right)C
               \\ & \quad + \Biggm[\eta\left( \beta_3\left( 1-\frac{C}{K_3} \right) + \mu_1 - \delta_3 \right)
               \left( \beta_2 - \delta_2 -\frac{\mu_2}{K_2}C \right) \\
                  &\quad -\frac{\beta_2}{K_2} \left( \delta_3 - \beta_3\left( 1-\frac{C}{K_3} \right) \right) \left( \frac{\mu_1}{K_1}C + \delta_1 \right)\Biggm]
               \\ & \quad \cdot
               \left( \delta_3 - \beta_3\left( 1-\frac{C}{K_3} \right) \right) \left( \frac{\mu_1}{K_1}C + \delta_1 \right) 
               ,
    \end{split}
  \end{equation}
  which correspond to the values of $C$ in a fixed point of
  \eqref{eq:lichenmodel}. 
  With $H(C), S(C)>0$ and Equations \eqref{eq:H(C)} and \eqref{eq:S(C)} follows
  that this is the case if
  \begin{equation}
    \label{eq:boundsforC}
    \check{C} \defl \frac{\beta_3-\delta_3}{\beta_3} K_3 < C < \frac{\beta_3-\delta_3+\mu_1}{\beta_3}  K_3 \defr \hat{C} .
  \end{equation}
  Thus, any fixed point of \eqref{eq:lichenmodel} must satisfy
  \eqref{eq:boundsforC}.
  In order to find roots lying in the interval $\opint{\check{C}, \hat{C}}$ we can
  evaluate $p$ at its boundaries, which yields
  \begin{equation*}
    p(\check{C}) = \eta^2\mu_1^2\mu_2 \frac{\beta_3-\delta_3}{\beta_3} K_3
  \end{equation*}
  and
  \begin{equation*}
    p(\hat{C}) = -\frac{\beta_2 \mu_1^2}{K_2} 
    \left( \delta_1 + \frac{\mu_1 K_3 (\beta_3-\delta_3+\mu_1)}{\beta_3 K_1}  \right)^2.
  \end{equation*}
  Note that $\check{C}$ is only a sensible lower bound for a root of $p$ leading
  to a relevant fixed point if $\check{C}>0$. 
  This is the case if and only if $\beta_3 > \delta_3$.
  We can furthermore see that $p(\hat{C}) \leq 0$ is always guaranteed and
  $p(\hat{C}) = 0$ if and only if
  \begin{equation}
    \label{eq:p(Chat)=0}
    \beta_3 - \delta_3 = -\frac{\beta_3 \delta_1 K_1}{\mu_1 K_3} - \mu_1,
  \end{equation}
  which can only be satisfied if $\beta_3<\delta_3$.
  This leads to the following three cases.

  \textit{Case 1}
  $\beta_3 > \delta_3$:
  We know that $p(\check{C})>0$ and since \eqref{eq:p(Chat)=0} cannot be
  satisfied, we have $p(\hat{C})<0$. 
  From the intermediate value theorem follows that at least one root of $p$ must
  lie in $\opint{\check{C}, \hat{C}}$.

  \textit{Case 2}
  $\beta_3 = \delta_3$:
  We find $\check{C}=p(\check{C})=0$ and we can get some information from
  considering the derivative of $p$ for $\beta_3=\delta_3$: 
  \begin{equation*}
    p'(0) = \eta^2\mu_1^2\mu_2 + \frac{\beta_3 \delta_1 \eta \mu_1}{K_3}(\beta_2-\delta_2)
  \end{equation*}
  Thus, as long as 
  $\beta_2-\delta_2 > -\frac{\eta \mu_1 \mu_2}{\beta_3 \delta_1} K_3$, 
  we have $p'(0)>0$ and therefore find some $\varepsilon>0$ such that
  $p(\varepsilon)>0$.
  Assuming that all parameters are positive, \eqref{eq:p(Chat)=0} cannot be satisfied
  and therefore we have $p(\hat{C})<0$. 
  From the intermediate value theorem follows again that there must be at least
  one root of $p$ in $\opint{0,\hat{C}}$. 

  \textit{Case 3}
  $\beta_3 < \delta_3$:
  It only makes sense to consider this case if $\mu_1 > \abs{\beta_3-\delta_3}$,
  because otherwise $\hat{C}\leq 0$, which implies that there is no interior
  fixed point according to \eqref{eq:boundsforC}.
  From \eqref{eq:p(Chat)=0} we can see that this condition also ensures
  $p(\hat{C})<0$, since 
  \begin{equation*}  
    \abs{\beta_3 - \delta_3} = -(\beta_3 - \delta_3) = \frac{\beta_3 \delta_1 K_1}{\mu_1 K_3} + \mu_1 \geq \mu_1 > \abs{\beta_3-\delta_3} 
  \end{equation*} 
  is clearly a contradiction.

  Since $\check{C}$ is not a sensible lower bound for $C$, we can instead evaluate
  $p$ at 0. 
  Thus, whenever
  \begin{equation*} 
    p(0)= - \delta_1 \eta (\beta_2-\delta_2)(\beta_3-\delta_3)(\beta_3-\delta_3+\mu_1) - \frac{\beta_2\delta_1^2}{K_2}(\beta_3-\delta_3)^2 > 0 
  \end{equation*}
  there will be a relevant fixed point, which again follows from the intermediate
  value theorem. 
  This is the case if and only if
  \begin{equation}
    \label{eq:interior-fp for beta3<delta3}
    \beta_2 - \delta_2 > -\frac{\beta_2\delta_1(\beta_3-\delta_3)}{\eta K_2 (\beta_3-\delta_3+\mu_1)}.
  \end{equation}
  We have shown that in these three cases there exist a (not necessarily unique)
  interior fixed point with $C\in \opint{\check{C}, \hat{C}[}$.
\end{proof}

\subsection{Derivation of System~\eqref{eq:reduction13_total}}
\label{apx:derivationsystem13}
The total population size of the host is $X=H+C$. 
On the slow manifold $C$ is given as a function of $H$ --- hence we find
\begin{equation*}
X = X(H)=H+C(H)= H + \frac{\alpha H}{\left(1 - \frac{H}{K_1}\right)},
\end{equation*}
which we need to solve for $H$. 
The solutions are the roots of the polynomial
\begin{equation*}
  H^2  - \left( K_1 ( 1 + \alpha ) + X \right) H + K_1 X
\end{equation*}
in $H$, which are given by
\begin{equation*}
  H_{\pm}(X) = \frac{1}{2} \left(K_1(1 + \alpha) + X 
  \pm 
  \sqrt{\left( K_1 (1 + \alpha) - X \right)^{2} + 4 \alpha K_1 X }\right).
\end{equation*}
From
\begin{align*}
  0 \leq \left( K_1(1 + \alpha)  - X \right)^{2} + 4 \alpha K_1 X 
  &= (K_1(1+\alpha) + X)^2 - 4 K_1 X \\
  &\leq (K_1(1+\alpha) + X)^2
\end{align*}
follows that both roots are real and $H_+(X)>K_1$.
Since we require $0\leq H < K_1$ and $X \geq 0$, only the solution $H_{-}(X)$ is
relevant, i.e.\ $H=H_{-}(X)$. 

Now we show that $H_{-}(X)$ is monotonically increasing with $X$.
We find
\begin{equation*}
  \frac{d}{dX} H_{-}(X) = \frac{1}{2}\left(1 - \frac{K_1 \left( \alpha - 1 \right) +
  X}{\sqrt{\left( K_1 \left( \alpha + 1 \right) + X \right)^{2} - 4 K_1 X}}\right) \geq 0
\end{equation*}
if and only if
\begin{equation*}
  \sqrt{\left( K_1 \left( \alpha + 1 \right) + X \right)^{2} - 4 K_1 X} 
  \geq
  K_1 \left( \alpha - 1 \right) + X .
\end{equation*}
If the RHS is negative, this holds because the expression in the square root is
nonnegative.
Otherwise, we can square both sides to obtain the equivalent inequality
\begin{equation*}
  \left( K_1 \left( \alpha + 1 \right) + X \right)^2 - 4 K_1 X - (K_1 \left( \alpha - 1 \right) + X)^2
  = 4 \alpha K_1^2 \geq 0,
\end{equation*}
which is always satisfied. 
Therefore, the autarkic population of the mycobiont increases with
its total population. 
However, we also find that the autarkic population is bounded from above, since
$\lim_{X\to\infty} H_{-}(X) = K_1$. 
This can be seen from
\begin{align*}
  \abs{K_1 - H_{-}(X)} &= K_1 - \frac{H_{-}(X) H_{+}(X)}{H_{+}(X)} \\
           &= K_1 - \frac{2 K_1 X}{K_1 (1 + \alpha ) + X + \sqrt{ (K_1(1+\alpha)
           + X)^2 - 4 K_1 X}} \\
           &\leq K_1 - \frac{K_1 X}{K_1 (1 + \alpha ) + X} \xrightarrow[X\to\infty]{} 0
\end{align*}

\section{Supplementary Material}
\label{apx:reductions}
A list of all TFPVs and the corresponding reductions for
System~\eqref{eq:lichenmodel} can be found as an ancillary file on arxiv.
The \texttt{Julia} script for finding and computing these reductions using
\texttt{TikhonovFenichelReductions.jl} can be found at
\url{https://github.com/jo-ap/TFR_ModellingMutualism}.

\printbibliography

\end{document}